\documentclass[amsmath,amssymb,aps,prx,superscriptaddress,12pt,floatfix]{revtex4-2}
\usepackage{setspace}
\usepackage{xcolor}
\usepackage{graphicx}
\usepackage{academicons}
\usepackage{orcidlink}
\usepackage{multirow}
\usepackage{chemformula}
\usepackage{booktabs}
\usepackage{array}
\usepackage{url}
\usepackage{makecell}
\usepackage{colortbl}
\usepackage{enumitem}
\usepackage{placeins}
\usepackage{hyperref}
\usepackage{xr-hyper}

\begin{document}

\title{Hierarchical high-throughput screening of alkaline-stable lithium-ion conductors combining machine learning and first-principles calculations}
\author{Zhuohan Li\,\orcidlink{0000-0001-5372-9450}$^\ast$}
\affiliation{Materials Sciences Division, Lawrence Berkeley National Laboratory, Berkeley, California 94720, United States}

\author{KyuJung Jun\,\orcidlink{0000-0003-1974-028X}$^\ast$}
\affiliation{Department of Materials Science and Engineering, University of California, Berkeley, California 94720, United States}
\thanks{Z.L and K.J. contributed equally to this work}

\author{Bowen Deng\,\orcidlink{0000-0003-4085-381X}}
\affiliation{Department of Materials Science and Engineering, University of California, Berkeley, California 94720, United States}

\author{Gerbrand Ceder \orcidlink{0000-0001-9275-3605}}
\email{Correspondence: gceder@berkeley.edu}
\affiliation{Materials Sciences Division, Lawrence Berkeley National Laboratory, Berkeley, California 94720, United States}
\affiliation{Department of Materials Science and Engineering, University of California, Berkeley, California 94720, United States}
\affiliation{Lead Contact}


\setstretch{2}

\begin{abstract}

Solid-state batteries require lithium-ion conductors that combine high ionic conductivity with stability under harsh electrochemical and chemical conditions. Here, we investigate the chemical factors governing the stability of NASICON-type and garnet-type Li-ion conductors in highly alkaline environments. This is particularly relevant to solid-state Li–air cells operated under humidified air where alkaline conditions arise due to the formation of LiOH discharge products. We implement a hierarchical high-throughput screening workflow that consists of a pre-screening step using a universal machine-learning interatomic potential and a more accurate DFT-based screening. This approach enables rapid evaluation of over 320,000 compositions, from which 209 alkaline-stable candidates are identified. We identify specific cation substitutions that improve alkaline stability in NASICON and garnet compounds and reveal the underlying mechanism. More importantly, we highlight design trade-offs that require careful composition optimization to simultaneously enhance synthesizability, operational stability, and Li-ion/electronic conductivities for practical humid Li-air battery applications.

\end{abstract}


\maketitle

\setstretch{2}

\section{Introduction}
\begin{figure*}[t]
\centering
\includegraphics[width=\linewidth]{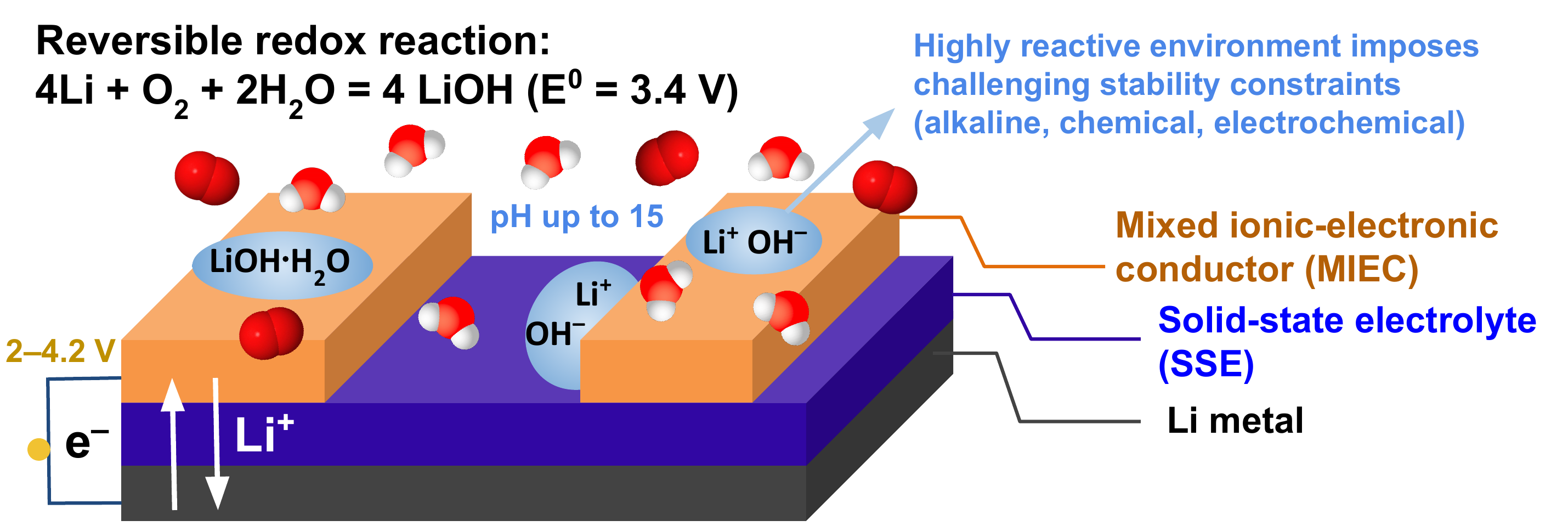}
\caption{\textbf{Schematic illustration of solid-state Li-\ch{O2} battery operated under humid environment} }
\label{fig:Li_air_cell} 
\end{figure*}

Li-ion-based energy storage technologies have emerged as a crucial solution to enable the transition to a net-zero carbon emission society. Despite remarkable advancements in current Li-ion batteries, the ever-growing energy demand necessitates the development of new technologies with improved capacity and cost efficiency. Solid-state batteries, in which the liquid electrolyte is replaced by solid-state electrolyte (SSE), are a promising next-generation system. Various types of SSE with different anion chemistries have been investigated so far, including oxide \cite{Thangadurai2014_garnet_review,Rossbach2018_NASICON_review,Stramare2003_perovskite_review,Kim2018_LTPO}, sulfide\cite{Zhou2021_argyrodite,Tao2022_LGPS,Kimura2023_LPS}, and halide-based\cite{Wang2022_halide_review,Kwak2022_halide_review} materials, each offering distinct advantages and disadvantages. Sulfide SSEs exhibit the highest Li-ion conductivities, but have the narrowest electrochemical stability window\cite{Richards2016_interface_stability,Zhu2016_interface,Xiao2020_interface}. Recent developments in halide conductors show the high potential of this chemistry as it may combine the high Li-ion conductivity and mechanical flexibility found in sulfides with the good oxidation stability of oxides \cite{Wang2022_halide_review}.

Water stability is a common issue among all three types of SSEs. Since water molecules are ubiquitous in the atmosphere, the instability against \ch{H2O} poses challenges in the handling and the long-period operation of solid-state batteries. For example, sulfide-based solid electrolytes such as $\ch{Li10GeP2S12}$\cite{Tao2022_LGPS}, argyrodite-type $\ch{Li6PS5Cl}$\cite{Zhou2021_argyrodite} are less practical due to their spontaneous reaction with $\ch{H2O}$ to form $\ch{H2S}$ gas \cite{Muramatsu2011_sulfide_reaction}. Chloride-based solid electrolytes such as $\ch{Li3YCl6}$ and $\ch{Li3InCl6}$ are also found to deteriorate rapidly upon absorption of water\cite{Feng2020_halide_air_stability}. Although the reactivity of oxides with \ch{H2O} is much lower, the \ch{Li+} in the oxides is subject to a topotactic exchange with \ch{H+} from water \cite{Cheng2018_LHX,Ye2021_garnet_LHX_review}, which may lead to a reduction of the Li-ion conductivity\cite{Sharafi2017_LHX_conductivity}. Moreover, this \ch{Li+}/\ch{H+} exchange promotes the formation of \ch{LiOH} on the surface of the oxide-based conductors \cite{Xu2024_LHX, Zou2020_LHX}. The highly hydrophilic nature of \ch{LiOH} \cite{Takeuchi2021_LiOH_hydration,Monnin2005_LiOH_pd} further facilitates water adsorption and formation of hydrated LiOH droplets \cite{Kim2023_operando}, creating a highly corrosive alkaline environment. The pH in these droplets can be as high as $\sim$15, determined by the saturation concentration of \ch{LiOH} in water \cite{Monnin2005_LiOH_pd}. 

A highly alkaline environment naturally arises in solid-state Li–O2 batteries when operated in humid air, where \ch{LiOH} forms as the discharge product as schematically illustrated in Figure \ref{fig:Li_air_cell}. In contrast to the current Li-ion batteries, where \ch{Li+} is inserted into and extracted from the cathode material, Li-\ch{O2} batteries store energy by forming various lithium oxides, such as peroxide, superoxide, and hydroxides, as the discharge product inside the cathode pores. The low mass density of these products makes the Li-\ch{O2} battery concept attractive for its uniquely high specific energy density \cite{Luntz2014_non_aqueous_Li_air,Liang2022_Li_air_review, Liu2020_nonaqueous_Li_air}. Conventional Li-\ch{O2} batteries are operated with dry \ch{O2} gas and a liquid electrolyte. Increased humidity facilitates the formation of \ch{LiOH} as the discharge product, replacing \ch{Li2O2} that forms under dry condition \cite{Guo2014_humidity,Liu2015_LiOH,Tan2016_humidity,Zhu2017_humidity}. The change in chemistry has several advantages. First of all, it can increase the discharge voltage from 2.96 V for \ch{Li2O2} to 3.4 V for \ch{LiOH}\cite{Gao2023_LiOH-based_review}. Second, the formation of \ch{LiOH} under humid air benefits the discharge capacity as the solution-mediated growth mechanism promotes \ch{LiOH} growth, in contrast to the insulating and growth-limited \ch{Li2O2}\cite{Schwenke2015_growth,Aetukuri2015_growth,Ma2018_humidity,Kim2022_carbon_free}. Finally, the higher ionic and electronic conductivity of LiOH also helps reduce the charge overpotential\cite{Liu2015_LiOH,Li2015_overpotential,Ma2018_humidity}. The potential parasitic reaction between \ch{H2O} in the air and the Li metal anode can be mitigated by replacing the liquid electrolyte with a SSE, which physically blocks water molecules from reaching the Li metal anode. In addition, a carbon-free cathode design by using a mixed-ionic electronic conductor (MIEC) as the porous cathode material can prevent the formation of undesirable \ch{Li2CO3} during the discharge, effectively reducing the charge overpotential \cite{Kim2022_carbon_free}.

To operate Li-\ch{O2} batteries under humid conditions, the surfaces of both the SSE and the MIEC cathode must be highly resistant to the highly alkaline and corrosive environment. In this work, we conduct high-throughput computational screening to identify alkaline-stable Li-ion conductors. Our focus is on improving the materials stability against the conditions encountered on the cathode side during battery operation, including high pH (alkaline stability), applied voltage (electrochemical stability), parasitic chemical reactions with chemical species in the environment (chemical stability). Our strategy is to perform compositional substitution into the NASICON and garnet crystal frameworks as they are known to exhibit high bulk Li-ion conductivity ($\sim$0.1-10 mS/cm) \cite{Rossbach2018_NASICON_review,Thangadurai2014_garnet_review}. The crystal structures of NASICON-type and garnet-type Li-ion conductors are depicted in Figure \ref{fig:crystal_pd}(a). 

To fully understand the tradeoffs between various properties across chemistry, we aim to explore a chemical space as broadly as possible with the aid of universal machine learning interatomic potential (uMLIP), which significantly accelerates the the evaluation of energies and properties. More specifically, we use the CHGNet MLIP model \cite{Deng2023_chgnet} to perform pre-screening over 320,000 different substitutions in NASICON and garnet frameworks. The pre-screened subsets of candidates are then further downselected through stability evaluations using density functional theory (DFT) calculations to achieve higher fidelity. Finally, we perform MLIP assisted molecular dynamics (MD) simulations on a subset of final candidate compounds using a fine-tuned CHGNet model to evaluate their Li-ion conductivities. Our hierarchical screening demonstrates a combination of MLIP, and DFT calculations to address materials design challenges.

\section{Screening Methodologies}
\subsection{Crystal frameworks and substitution chemical space}

\begin{figure*}[t]
\centering
\includegraphics[width=\linewidth]{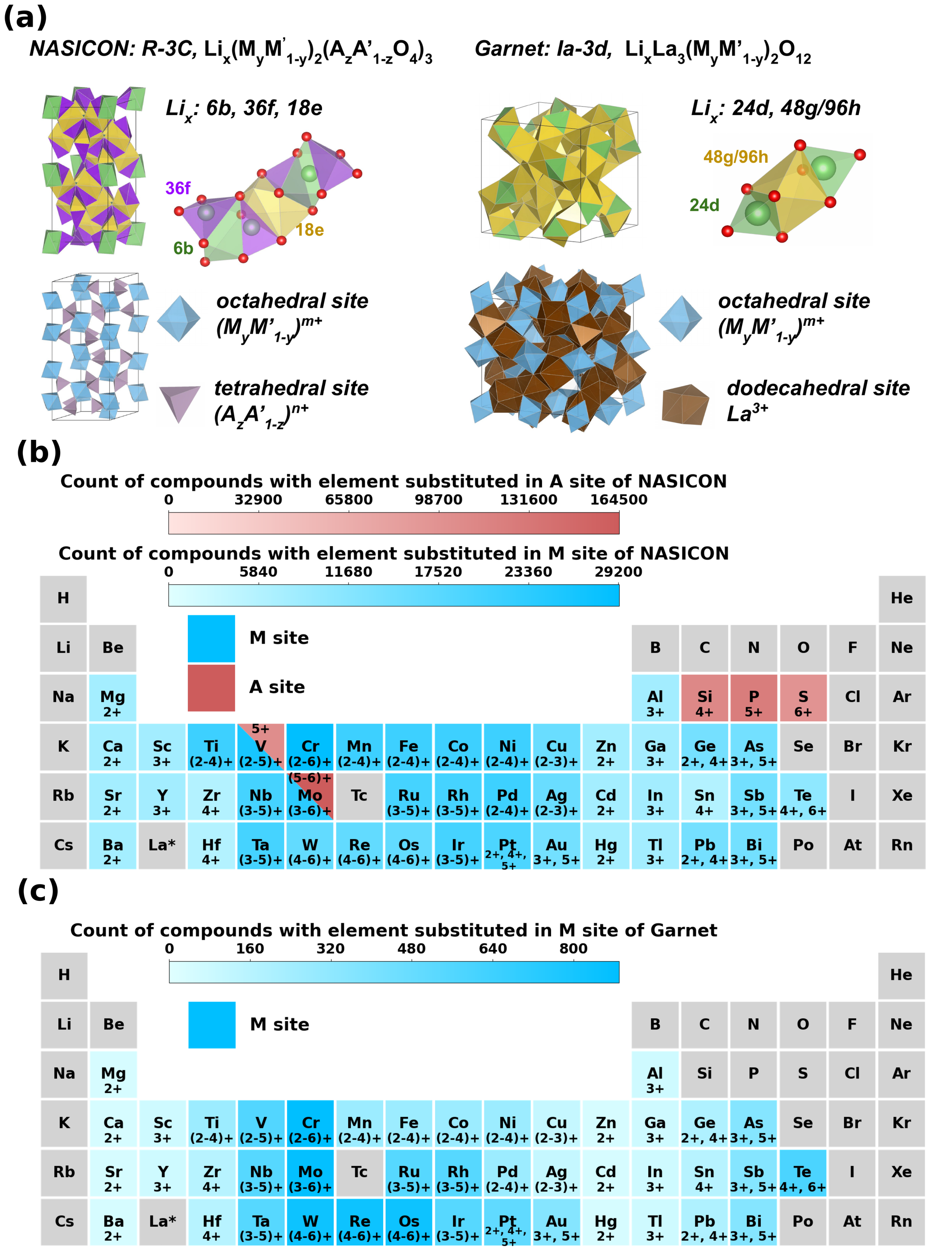}
\caption{\textbf{Crystal frameworks and substitution chemical space for screening} (a) Crystal frameworks and substitution sites in the Li-ion conductors. The number of initial candidate compounds substituted with different elements across the periodic table in A site (blue) or M site (red) of (b) NASICON and (c) garnet.}
\label{fig:crystal_pd} 
\end{figure*}

For NASICON compounds, we adopt the crystal structures with R$\overline{3}$C symmetry and with the chemical formula Li$_x$(M$_y$M'$_{1-y}$)$_2$(A$_z$A'$_{1-z}$O$_4$)$_3$. The cations A and A' occupy the tetrahedral sites that are coordinated by four oxygen atoms forming the polyanion group, and M and M' occupy octahedral sites that are corner-sharing with the six surrounding tetrahedral sites. M and M' represent a pair of transition metal cations, selected based on commonly observed oxidation states. We systematically enumerate substitutional compositions by varying the Li content $x$ between 1.0 to 3.0 with increments of 0.5 and by considering all possible M$_y$M'$_{1-y}$ cation pairs within the transition metal species set. The average valence state of M$_y$M'$_{1-y}$ is constrained between 2+ and 6+, incremented by  0.25. Overall charge neutrality is achieved by combining different mixtures of polyanion groups with diverse formal charges, including (PO$_4$)$^{3-}$, (SO$_4$)$^{2-}$, (SiO$_4$)$^{4}$, (VO$_4$)$^{3-}$, and (MoO$_4$)$^{2-/3-}$. To include both SSEs and MIECs, we sample a vast range of substitutions of the M site, including both redox-inactive single-valent and multi-valent cations that can undergo redox, resulting in 313,217 compositions in total. The substituted elements and their allowed valence states for NASICON are presented in the periodic table in Figure \ref{fig:crystal_pd}(b). The color intensity indicates the number of compounds in our set that contain this element either on the M site (blue) or on the A site (red).

For garnet compounds, we adopt the crystal structures with Ia$\overline{3}$d symmetry with the chemical formula Li$_x$La$_3$(M$_y$M'$_{1-y}$)$_2$O$_{12}$, where $x$ ranges from 3.0 to 7.0 in increments of 0.5. Similar to NASICON, M and M' denote transition metal cations in different oxidation states, with the average oxidation state of M$_y$M'$_{1-y}$ constrained between 2+ and 6+ in steps of 0.25. We only substitute the octahedral M site while fixing La$^{3+}$ in the dodecahedral site. A total of 6,823 La-based garnet candidates are generated, with  their elemental distribution shown in  Figure \ref{fig:crystal_pd}(c). Although the initial number of garnet candidates is considerably smaller than that of NASICONs, we show below that the pass rate of garnet compounds while screening is substantially higher, yielding a comparable number of final candidates: 121 SSEs and 7 MIECs for garnet, versus 74 SSEs and 7 MIECs for NASICON. This high pass rate for the garnet compounds suggests a generally greater alkaline stability of garnets relative to NASICONs, as discussed in later sections.

\subsection{Stability metrics}

\begin{figure*}[t]
\centering
\includegraphics[width=\linewidth]{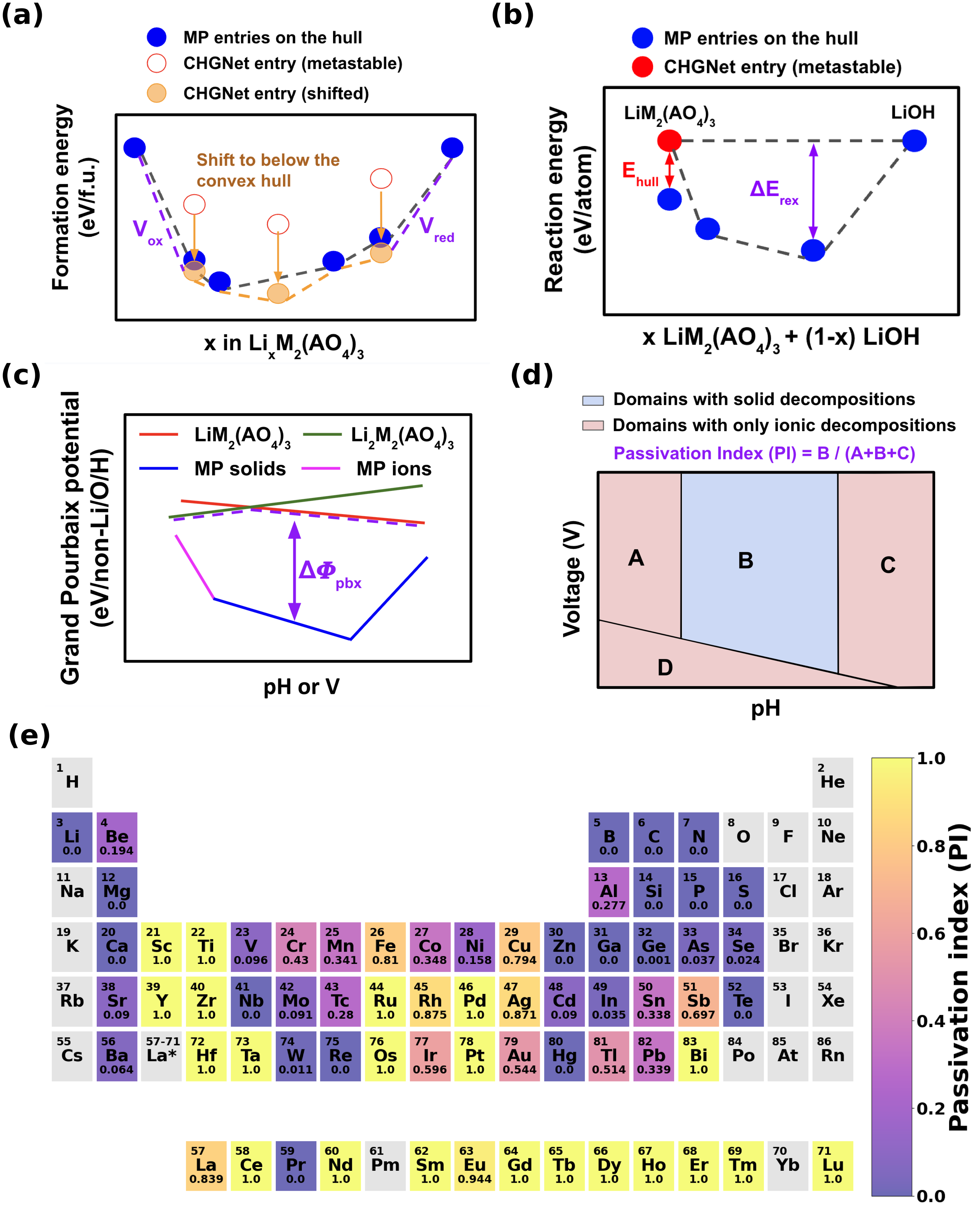}
\caption{\textbf{Schematic plot of stability metrics} (a) $V_{ox}$ and $V_{red}$ for electrochemical stability. (b) $\Delta E_{rxn}$ for chemical stability. (c) $\Delta\phi_{pbx}$ and (d) passivation index (PI) for alkaline stability. (e) Elemental PI across the periodic table. In Figures (a-c), NASICON compounds \ch{Li$_x$M2(AO4)3} are assumed for demonstration purpose, but the stability metrics are similarly defined for garnet compounds \ch{Li$_x$La3M2O12} as well.}
\label{fig:stability_metric} 
\end{figure*}

In this section, we explain how we quantify the potential synthesizability of the candidate materials, as well as their stability against the harsh environments in humid Li-air systems. To exclude highly unstable compounds that are unlikely to be synthesized, we calculate the energy above the convex hull ($E_{hull}$) for each candidate. Although reasonably low $E_{hull}$ is not a sufficient condition for the successful synthesis of a hypothetical material, it serves as an efficient and widely used metric for ruling out highly unstable materials and serves as a rough proxy of the synthesizability of a material \cite{Ouyang2021_NASICON, Sun2016_synthesis, Aykol2018_thermo_limit}.

In addition to $E_{hull}$, we assess the driving force for three distinct parasitic reactions that can degrade the material as illustrated in Figures \ref{fig:stability_metric}(a-d). We evaluate the electrochemical stability against the applied voltages, using the previously developed approach \cite{Mo2012_voltage,Richards2016_interface_stability,Zhu2015_voltage}, and calculate the electrochemical stability window for each candidate. The upper and the lower limits correspond to the oxidation ($V_{ox}$) and the reduction ($V_{red}$) limits, respectively (Figure \ref{fig:stability_metric} (a)). We apply two modifications to the general approach \cite{Richards2016_interface_stability,Zhu2015_voltage} for the estimation of the electrochemical window. (1) In theory a metastable material ($E_{hull}$ $>$ 0 meV/atom) has no electrochemically stable window as thermodynamics would predict its decomposition to its competing phases even without voltage applied. However, metastable materials often exhibit certain stability against oxidation and reduction, which can be due to various reasons, including slow kinetics of decomposition at room temperature, thermal stabilization, and formation of passivating decomposition products on the surface, etc. To obtain an optimistic estimation of the electrochemical stability window of a metastable compound, we vertically shift the energy of the metastable target compound, breaking the convex hull at the corresponding composition (either at the stable decomposition on the hull or on the tie line between the nearest neighbor stable compositions) by a small amount (10 meV/atom for CHGNet pre-screening step, and 1 meV/atom for DFT-screening step). The resulting electrochemical window represents the upper (lower) limits of the true thermodynamic oxidation (reduction) limits. (2) Redox-active compounds can undergo topotactic lithium insertion and extraction reactions instead of decomposition reactions \cite{Jun2023_Li6CoO4}. As the topotactic reaction maintains the host crystal framework, we do not consider this a "breakdown" of a material. Thus, we assume that the topotactic Li insertion and extraction is kinetically accessible for topotactically related compositions. Therefore, the oxidation (or reduction) limit of a redox-active compound is defined as the maximum (or minimum) of all oxidation (or reduction) limits among all topotactically related compositions with different Li contents.

During the battery operation, the surfaces of SSE and MIEC cathode are exposed to \ch{H2O} and \ch{LiOH}. \ch{H2O} is naturally present due to the exposure of the cathode to humid air, while $\ch{LiOH}$ is a discharge product on the surfaces of SSE or MIEC cathode. We compute the driving force for the chemical reaction between our candidate compounds and \ch{H2O} or \ch{LiOH}. As shown in Figure \ref{fig:screening_scheme}(c), the reaction product and the corresponding reaction energy depend on the ratio between the two reactants. Similar to concepts used in solid-state reaction theory \cite{Szymanski2024_synthesis}, we use the most negative reaction energy per reactant atom as an indicator of the driving force for our candidate materials to chemically react with \ch{H2O} or \ch{LiOH}, referring to it as $\Delta E_{rxn}$. Since $\Delta E_{rxn}$ is referenced to the reactants, a spontaneous reaction will have a negative value of $\Delta E_{rxn}$. 

The formation of $\ch{LiOH}$ as the discharge product in the humid environment inevitably creates a high pH environment due to the hydrophilic nature of $\ch{LiOH}$. Therefore, the Li-ion conductors also need to be highly resistant to the high pH conditions, which can cause the dissolution of many species. In our screening, the alkaline stability is evaluated by constructing Pourbaix diagrams for each candidate compound \cite{Persson2012_pbx}. A Pourbaix diagram shows whether the target compound is stable, or if not, which set of thermodynamically most stable decomposition products (either solid, ions in solution, or the combination of both) form at a given pH and voltage. In the Pourbaix diagram, the stability of the target compound at a given pH and voltage is quantified by the so-called Pourbaix potential ($\psi_{pbx}$) \cite{Sun2019_pourbaix,Wang2024_pbx}. The difference in the Pourbaix potential between the target compound and the most stable decomposition phases (either solid phases or ionic solution species) is called the Pourbaix decomposition energy, which measures the thermodynamic driving force to decompose in a given pH and voltage condition \cite{Singh2017_pbx_metastab}. Different to $\Delta E_{rxn}$ vs. LiOH, where only solid phases are taken into consideration, the Pourbaix decomposition energy is evaluated including both solid phases and ionic solution species within a given chemical space, and thus, it better describes the stability of materials in a solution environment.

Conventionally, the Pourbaix diagram assumes that the system is open to oxygen, protons, and electrons, as these species can be delivered or absorbed by the oxidation/reduction reactions in water. As a result, the Pourbaix potential has a unit of energy per non-O/H atoms (eV/non-O/H). Here, in order to simulate the alkaline stability of both SSE and MIEC compounds during the battery operation with varying lithium chemical potential, we further assume that the system is also open to \ch{Li+} exchange, in which case the Pourbaix potential has a unit of energy per non-Li/O/H atoms (eV/non-Li/O/H). To differentiate it from the conventionally calculated Pourbaix potential, we refer to the new potential as the grand Pourbaix potential ($\phi_{pbx}$). This newly-defined grand Pourbaix potential is particularly useful for estimating the alkaline stability of compounds where \ch{Li+} can be topotactically inserted or extracted from the host structure as the voltage varies during charge and discharge. As demonstrated in Figure \ref{fig:screening_scheme}(d), the grand Pourbaix potential of these topotactically related compositions can be evaluated simultaneously in the same Pourbaix diagram. At each pH and voltage condition, the Li content of a given non-Li composition is determined as the one with the lowest $\phi_{pbx}$. Following the convention in Ref \cite{Singh2017_pbx_metastab}, the value of $\Delta\phi_{pbx}$ is referenced to the most stable phases, and therefore, any metastable compound will have positive values of $\Delta\phi_{pbx}$. We use the maximum of the grand Pourbaix decomposition energy (max $\Delta\phi_{pbx}$) within the battery operating window (pH = $12-15$ and voltage = $2.0-4.2$ V vs. Li/\ch{Li+}) as an indicator for alkaline stability. The candidate material with lower max $\Delta\phi_{pbx}$ is considered to be more resistant to the corrosive alkaline environment.

That being said, a material with a high max $\Delta\phi_{pbx}$ value does not necessarily imply the material always severely degrades. It was previously demonstrated that the degradation of a material can be potentially mitigated if the decomposition products can act as an effective corrosion-resistive solid passivation layer \cite{Singh2017_pbx_metastab}. Passivation is controlled by a series of chemical and mechanical effects on the surface and cannot be exactly predicted with any models \cite{Xu2000_PB_ratio_alloy,Barthel2021_limitation_pbx, Wang2023_passivation_review, Walters2021_Cu_corrosion}. In this work, we therefore estimate the passivation capability of candidate compounds by a passivation index (PI). PI is defined as the fractional area of the domains in the Pourbaix diagram where at least one solid decomposition product forms within the battery operation window (pH = $12-15$ and voltage = $2.0-4.2$ V vs. Li/\ch{Li+}). A higher PI value indicates a higher chance that the surface can be covered by a solid passivation layer and be protected from further dissolution. In practice, a solid product may not always function as a stable passivation layer \cite{Xu2000_P-B_ratio,Macdonald2007_passivation_theory}. Therefore, PI can be seen as the best-case estimation of the passivation capability of a given composition. 

The PI of each element's individual Pourbaix diagram is shown in Figure \ref{fig:stability_metric} (e) and provides useful information about the passivation capability of each element. As can be seen, there is a large variation in the elemental PI across the periodic table. Some elements, such as early transition metals (TMs) and the lanthanide series, have very high PI close to 1.0, whereas others, including alkaline earth metals and first-row 3d TMs, exhibit negligible passivation capability with PI values approaching 0.0. According to our definition, the PI for a compound will always be equal or greater than that of the consisting element in the compound with the highest PI. This is because different elements in a compound can form different solid decomposition products at different pH and voltage conditions. It is also possible that multiple elements form a compound-type decomposition product. Therefore, elements with low elemental PI can be co-substituted with those having high elemental PI to increase the overall PI of the compound. Our model implicitly assumes that, in the worst-case scenario, surface atoms with low elemental PI dissolve into the solution, whereas the remaining atoms with high elemental PI will accumulate and form a passivation layer on the surface. Elemental La exhibits a high PI value of 0.84, and other elements in the lanthanide series show even higher values. This suggests that lanthanides generally offer strong passivation capabilities and are unlikely to be the limiting factor for passivation in garnet materials. 

\subsection{Hierarchical high-throughput screening scheme}
\label{subsec:screen_scheme}
\begin{figure*}[t]
\centering
\includegraphics[width=0.7\linewidth]{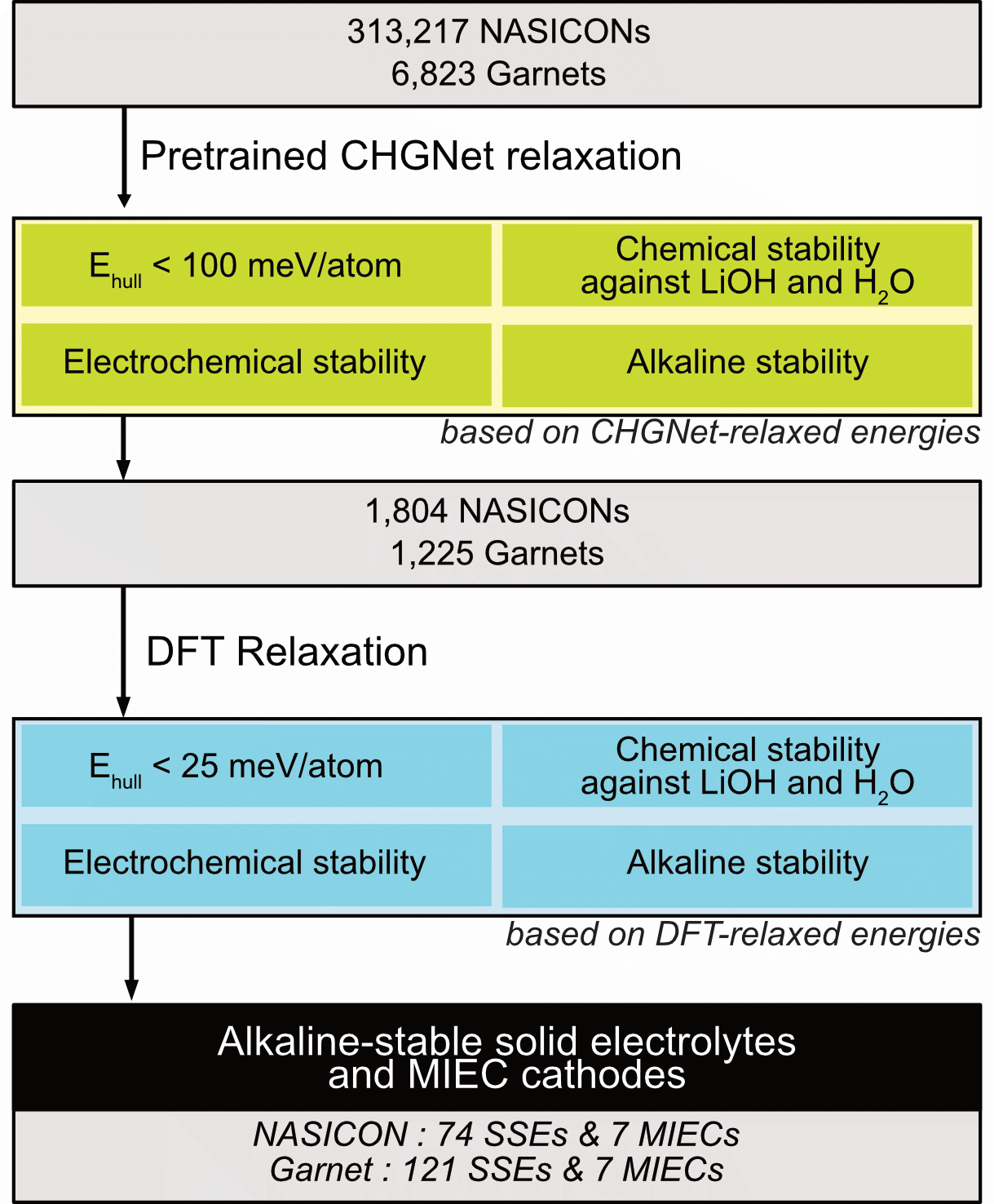}
\caption{\textbf{2-step hierarchical screening scheme combining MLIP and DFT calculations}} We use pretrained CHGNet to relax all initial candidate compounds in the pre-screening step. DFT relaxations are performed on the subset of candidates that pass the pre-screening step.
\label{fig:screening_scheme} 
\end{figure*}

Figure \ref{fig:screening_scheme} illustrates the hierarchical screening scheme that combines the use of the CHGNet model \cite{Deng2023_chgnet} and DFT. CHGNet is a universal machine-learning interatomic potential (uMLIP) model pretrained on the MPtrj dataset, which consists of over 1.5 million inorganic materials calculated by the Materials Project (MP) team with DFT \cite{Jain2018_MP,Horton2025_MP_new_review}. The CHGNet model has previously been shown to have reasonable accuracy in studying Li-ion conductors \cite{Jun2024_nitride, Zhong_2024_LZC}. We apply the CHGNet model for a fast pre-screening (see Figure S\ref{fig:chgnet_error} for the benchmark test on the CHGNet-relaxed energy), followed by higher fidelity DFT calculations on a selected subset of compounds. Starting with 313,217 and 6,823 candidates with NASICON and garnet structures,  we perform structure relaxation using the pretrained CHGNet. In the pre-screening step, energy above hull is computed by comparing the CHGNet-relaxed energy with the convex hull constructed from all competing phases available in the MP. Compounds with $E_{hull} < 100$ meV/atom are selected for further evaluation of their chemical, electrochemical, and alkaline stabilities (see Table S\ref{tab:screen_criteria_NASICON} and Table S\ref{tab:screen_criteria_garnet} in the Supplemental Information for detailed screening criteria). This pre-screening step yields 1,804 NASICON and 1,225 garnet candidate compounds. In the subsequent DFT-screening step, we re-relax the pre-screened candidates with DFT calculations, starting from the CHGNet-relaxed structures. The energy above hull is reevaluated using the DFT-relaxed energy, and the candidates are further downselected with a tighter criterion $E_{hull} < 25$ meV/atom. For the compounds that meet this criterion, we recalculate chemical, electrochemical, and alkaline stabilities using the DFT-relaxed energies. The screening criteria of chemical, electrochemical, and alkaline stabilities for NASICON and garnet compounds differ slightly between the CHGNet pre-screening and DFT downselection steps. The rationales behind the screening criteria are detailed in the Section S\ref{sec:rationale} in the Supplemental Information. Our hierarchical screening ultimately identifies 81 NASICON (74 SSEs and 7 MIECs) and 128 garnet (121 SSEs and 7 MIECs) compounds.

\section{Results}
\subsection{Distributions of stability metrics among all initial candidate compounds}

\begin{figure*}[t]
\centering
\includegraphics[width=\linewidth]{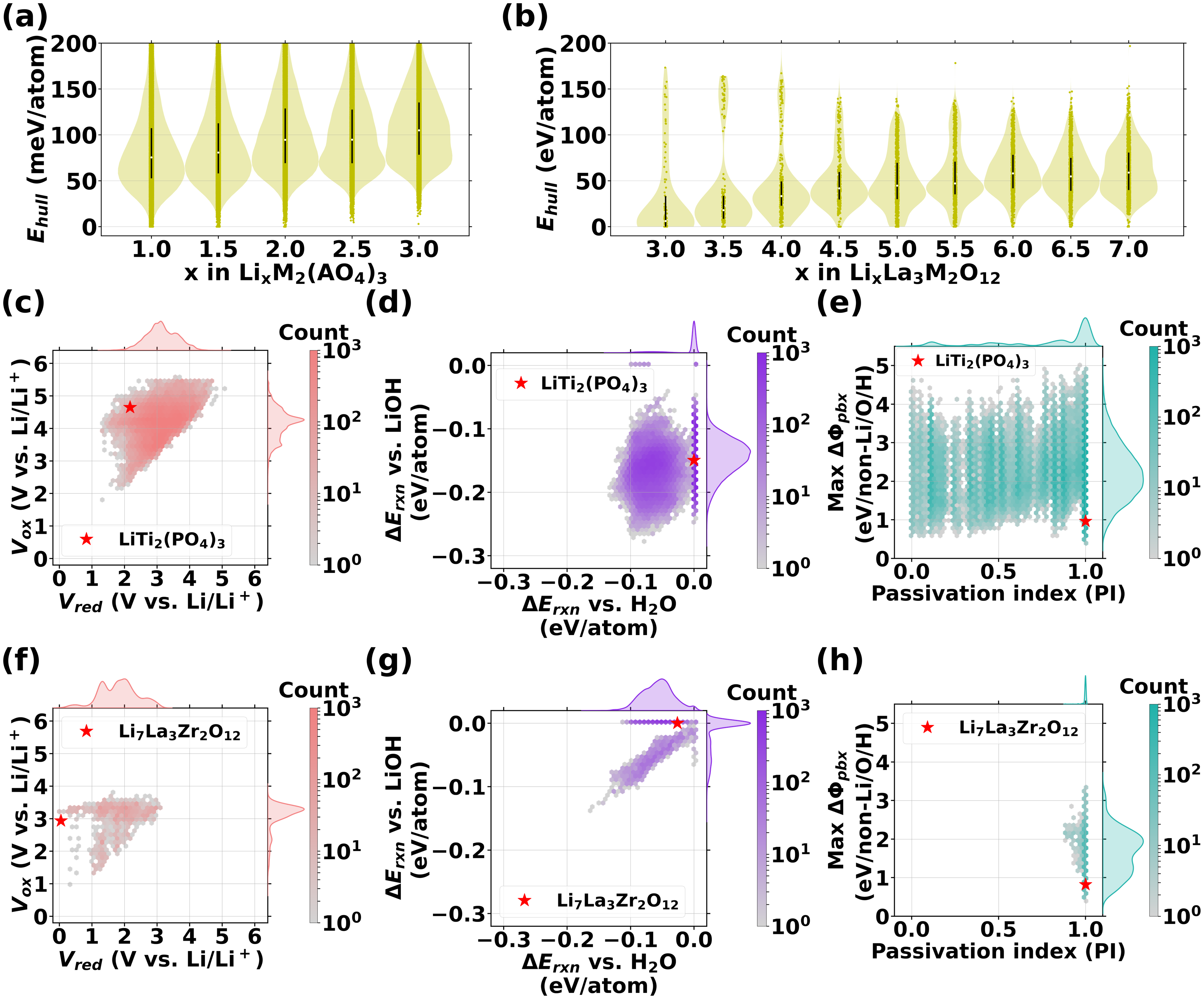}
\caption{\textbf{Distributions of stability metrics based on CHGNet-relaxed energies} Synthesizability ($E_{hull}$) for (a) NASICON and (b) garnet. (c) electrochemical stability ($V_{ox}$ and $V_{red}$), (d) chemical stability ($\Delta E_{rxn}$ vs. \ch{H2O} or \ch{LiOH}), and (e) alkaline stability (max $\Delta\phi_{pbx}$ and PI) of NASICONs. The corresponding distributions for garnets are shown in (f-h). In (c-h), only the compounds with $E_{hull}$ $<$ 100 meV/atom are included. The data points for \ch{LiTi2(PO4)3} and \ch{Li7La3Zr2O12} shown in (c-h) are obtained by using their DFT-relaxed energies for higher accuracy.}
\label{fig:chgnet_screening} 
\end{figure*}

The distributions of the stability metrics based on CHGNet-relaxed energies for both NASICON and garnet compounds are presented in Figure \ref{fig:chgnet_screening}. In Figures \ref{fig:chgnet_screening} (a) and (b), the distribution of $E_{hull}$ values are shown as a function of Li content per formula unit ($x$) for NASICON and garnet, respectively. Our screening results show that the distributions of $E_{hull}$ for garnets has overall lower values than those of NASICONs. Such trend persists even when only phosphate-based NASICON compounds are considered (see Supplemental Information Figure S\ref{fig:Ehull_phosphate_nasicon}). These results imply that La-based garnets may intrinsically be easier to substitute than NASICON compounds. We also observe that the trend that $E_{hull}$ increases with lithium content $x$ for both crystal frameworks, indicating a decrease of stability as more Li is stuffed into the host structure. Typically, increasing Li content beyond the baseline content ($x = 1$ for phosphate-based NASICON and $x = 3$ for garnet) enhances Li-ion conductivity by creating activated local environments that diffuse with low barrier without getting trapped or dissipated\cite{Xiao2021_NASICON_garnet}. For phosphate-based NASICON, the optimal conductivity was found to be in the range of 1.0 $<$ $x$ $<$ 2.0 \cite{Rossbach2018_NASICON_review}, while the range 6.0 $<$ $x$ $<$ 7.0 is optimized for garnet \cite{Thangadurai2014_garnet_review}. Given that the $E_{hull}$ range for experimentally synthesized oxide compounds can often reach up to $\sim$60 meV/atom \cite{Sun2016_thermo_scale}, our screening results reveal that there are potentially many undiscovered Li-stuffed NASICON and garnet compounds that are synthesizable and promising as new Li-ion conductors.

For all compounds with CHGNet-calculated $E_{hull} < 100$ meV/atom, we show the distribution of electrochemical stability, characterized by $V_{ox}$ and $V_{red}$ in Figure \ref{fig:chgnet_screening} (c) and (f) for NASICONs and garnets, respectively. Our analysis shows that NASICONs have higher values for both $V_{ox}$ and $V_{red}$ than garnets, suggesting that NASICONs are more stable against oxidation, whereas garnets are superior in terms of the stability against reduction, consistent with previous theoretical and experimental studies \cite{Richards2016_interface_stability, Han2016_lgps_llzo}. The trend in $V_{red}$ can be understood by the inductive effect of the polyanion groups in NASICONs \cite{Masquelier2013_polyanion_review}. The strong covalent bonds within the polyanion unit lowers the covalency of the bonds between the octahedral-site cation and the surrounding oxygen anions, leading to an increase in the potential of the redox couple vs. Li/\ch{Li+}. Similarly, the oxidation of the oxygen anion, which usually is the breakdown mechanisms for oxide SSE at high voltages is moved to higher voltage by its covalent bonding in the polyanion groups \cite{Masquelier2013_polyanion_review}. As our target voltage range is 2.0 - 4.2 V vs. Li/\ch{Li+}, the compositional optimization should aim to lower the reduction limit for NASICON and raise the oxidation limit for garnet.

The chemical stability against side reactions is quantified by the reaction energy ( $\Delta E_{rxn}$) between the given Li-ion conductors and chemical species prevalent in the environment, namely \ch{H2O} and \ch{LiOH}. As can be seen in Figure \ref{fig:chgnet_screening} (d) and (g), garnet compounds tend to be more reactive against \ch{H2O} with a $\Delta E_{rxn}$ magnitude peaking at $\sim$-0.05 eV/atom, while NASICON compounds show near zero reactivity against \ch{H2O}. In contrast, the majority of NASICON compounds easily react with the discharge product \ch{LiOH} with  $\Delta E_{rxn}$ peaking at $\sim$-0.14 eV/atom, whereas garnets are much more stable against \ch{LiOH}. The high reactivity between NASICON and \ch{LiOH} poses concerns for using NASICON-type compounds in a high pH environment. Indeed, experimental observation of the degradation of NASICON-type conductors in alkaline solutions has been previously reported \cite{Lam2024_LTGP_high_pH,Hasegawa2009_alkaline_NASICON,Shimonishi2011_alkaline_NASICON}.

The distribution of  $\Delta\phi_{pbx}$ and PI values are plotted in Figures \ref{fig:chgnet_screening} (e) and (h). The results shown in Figure \ref{fig:chgnet_screening} indicate that most of the NASICON and garnet compounds have good passivation capabilities with the peak of the PI distribution near 1.0. We also find that the distribution of PI is much broader for NASICONs compared to garnets, indicating that the high passivation capability is more consistent among different garnet compounds, whereas it is more chemistry-dependent for NASICON. The driving force for decomposition in an alkaline medium peaks near max $\Delta\phi_{pbx}$ = $\sim$2.0 eV/non-Li/O/H for both crystal frameworks. However, the distribution for garnet is slightly narrower and skews more towards lower max $\Delta\phi_{pbx}$ values compared to that of NASICON. Both the narrower PI distribution and the lower max $\Delta\phi_{pbx}$ values indicate that garnet compounds would generally be less prone to degradation in an alkaline environment compared to NASICON.

For comparison, the stability metrics for the prototype compounds of NASICON (\ch{LiTi2(PO4)3}, LTPO) and garnet (\ch{Li7La3Zr2O12}, LLZO) are also calculated using DFT, which are labeled as red stars in the Figures \ref{fig:chgnet_screening} (c-h). These prototype compounds show contrasting behavior in terms of electrochemical and chemical stabilities: while LTPO is more stable against oxidation with higher $V_{ox}$ of 4.6 V, LLZO has $V_{red}$ close to zero, indicating the significant stability of LLZO against Li metal \cite{Richards2016_interface_stability}; LTPO is more reactive with \ch{LiOH} with $\Delta$ E$_{rex}$ = -0.15 eV/atom, while it remains stable against \ch{H2O} with zero reaction energy. Conversely, LLZO spontaneously reacts with \ch{H2O} with $\Delta$ E$_{rex}$ = -0.03 eV/atom, but it is stable against LiOH with zero reaction energy. These results on electrochemical and chemical stabilities align with the trend observed across all NASICON and garnet compounds. As for alkaline stability, both LTPO and LLZO show similar passivation behavior with PI = 1.0 for both compounds. Also, both LTPO and LLZO have low max $\Delta\phi_{pbx}$ values of 1.0 and 0.8 eV/non-Li/O/H, respectively. The distributions shown in Figure \ref{fig:chgnet_screening} indicate the presence of numerous compounds that may potentially have superior stability compared to LTPO and LLZO, characterized by less negative $\Delta$ E$_{rex}$ values, higher oxidation and lower reduction limits, and less positive max $\Delta\phi_{pbx}$ values. These NASICON and garnet compounds are first filtered based on the screening criteria for CHGNet pre-screening step, and further downselected using DFT-relaxed energies (see Tables S\ref{tab:screen_criteria_NASICON} and S\ref{tab:screen_criteria_garnet} in the Supplemental Information), yielding 81 NASICON and 128 garnet compounds as the final candidates. 

\FloatBarrier
\subsection{Final list of compounds as candidate SSE and MIEC materials}

\begin{figure*}[t]
\centering
\includegraphics[width=\linewidth]{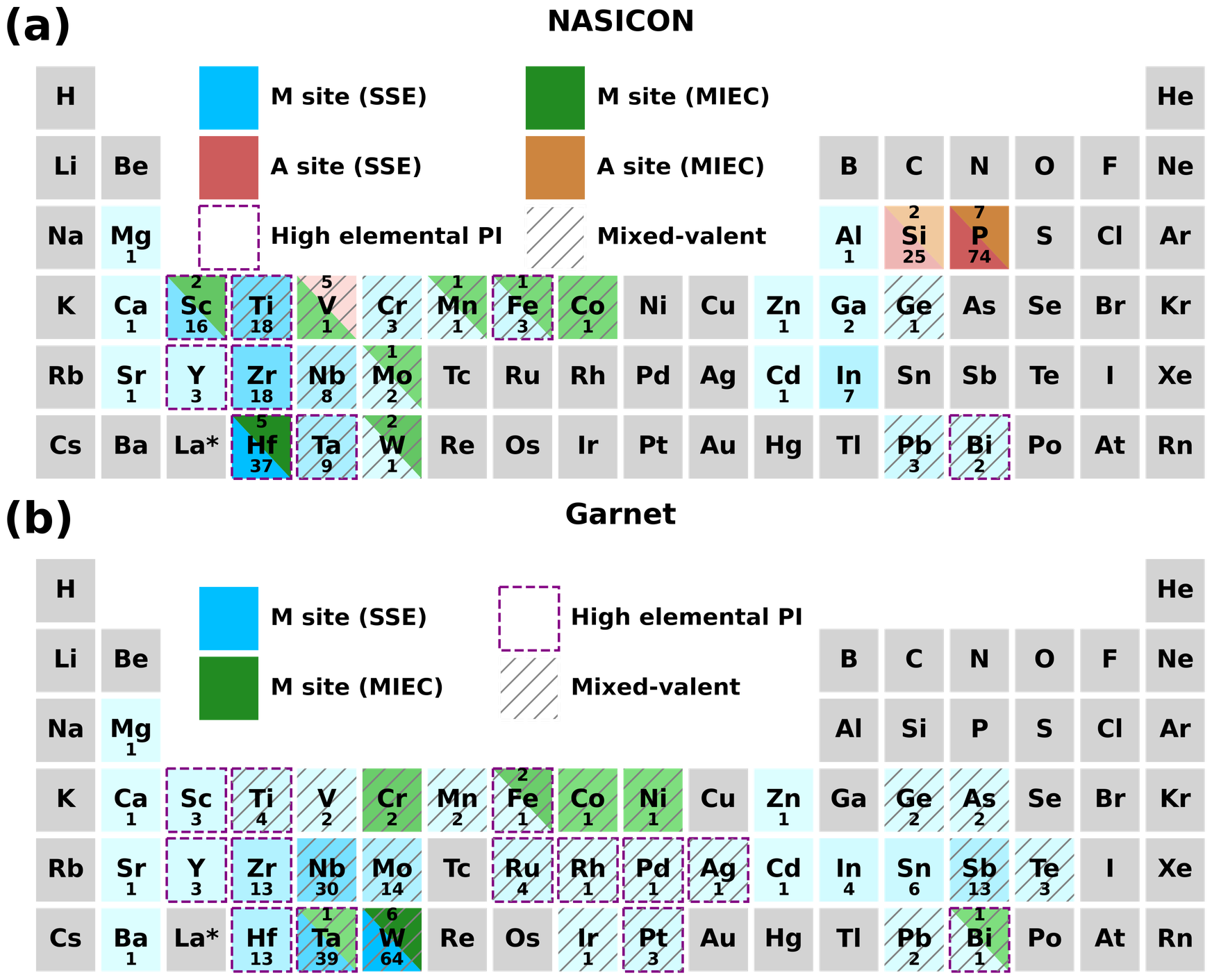}
\caption{\textbf{Elements selected in the final candidate compounds} (a) Substituted elements in M and A sites among 74 SSE and 7 MIEC candidates with NASICON framework. (b) 121 SSE and 7 MIEC candidates with garnet framework. M-site substitutions are shown in blue and green for SSE and MIEC compounds, respectively.  A-site substitutions are shown in red and brown for SSE and MIEC compounds, respectively. The elements with high elemental PI (PI $>$ 0.8) are highlighted with dashed purple lines around the boxes. The mixed-valent elements are highlighted with green slashes inside the boxes. The number in each elemental box represents the number of final candidate compounds that contain the corresponding element. All topotactically related compositions that only differ by Li content are grouped and counted as one MIEC compound. }
\label{fig:final_candidates} 
\end{figure*}

\begin{table*}[t]
\caption{\label{tab:table3} \textbf{A subset of the final screened NASICON candidates} 
All identified NASICON MIECs and 10 out of the 74 final NASICON SSE candidates are tabulated here. Only the lowest $E_{hull}$ value among different Li contents is listed for each MIEC compound with varying Li-content, where the corresponding Li content is indicated in the parenthesis. The parenthesis in the Li intercalation voltage indicates the range of lithium content at which its corresponding voltage lies outside the battery operating window ($2.0-4.2$ V vs. Li/\ch{Li+}). The Li-ion conductivity at 300 K is evaluated by performing molecular dynamics simulations using a fine-tuned CHGNet model. The numbers in brackets are the upper and lower limit estimation of the Li-ion conductivity, and the error for activation energy represents the standard deviation value \cite{He2018_MD_error}. For each MIEC compound, Li-ion conductivity is evaluated only at the composition specified in parenthesis. This composition meets all DFT-screening criteria and has a Li content x closest to 1.5, which has been experimentally shown to yield higher Li-ion conductivity \cite{Rossbach2018_NASICON_review}.
}
{
\fontsize{8}{10}\selectfont
\renewcommand{\arraystretch}{2.0}
\rowcolors{2}{gray!20}{white}
\begin{ruledtabular}
\begin{tabular}{ccc>{\centering\arraybackslash}p{2.4cm}>{\centering\arraybackslash}p{2.5cm}>{\centering\arraybackslash}p{2.3cm}c} 
 Compound&\makecell{$E_{hull}$\\(meV/atom)}&\makecell{Max $\Delta\phi_{pbx}$\\(eV/non-Li/O/H)}
&\makecell{Li intercalation\\voltage (V)}&\makecell{Li-ion conductivity\\at 300 K (mS/cm)}&\makecell{Activation\\energy (eV)}&Category\\ \hline
 \ch{LiTi2(PO4)3}&0&0.963&(Li$_1$-Li$_3$: 1.5)&0.07 (Exp.)\footnotemark[1]&0.47 (Exp.)\footnotemark[1]&\makecell{Baseline\\SSE}\\
 \ch{Li3Sc2(PO4)3}&14&0.497&None&2e-4, [6e-5, 8e-4]&$0.560\pm0.031$&SSE\\
 \ch{LiTa2Si2PO12}&16&0.579&None&2e-4, [4e-5, 7e-4]&$0.560\pm0.034$&SSE\\
 \ch{Li2Hf_{1.5}Sc_{0.5}Si_{0.5}P_{2.5}O12}&17&0.580&None&1.4, [0.9, 2.2]&$0.288\pm0.010$&SSE\\
 \ch{Li_{1.5}Hf2Si_{0.5}P_{2.5}O12}&2&0.596&None&3.2,[2.1, 4.9]&$0.243\pm0.010$&SSE\\
 \ch{LiTa_{5/3}Nb_{1/3}Si2PO12}&19&0.602&None&4e-5,[9e-6, 2e-4]&$0.605\pm0.038$&SSE\\
 \ch{Li_{2.5}Hf_{0.5}Sc_{1.5}(PO4)3}&22&0.603&None&3.2, [2.2, 4.8]&$0.260\pm0.009$&SSE\\
 \ch{Li_{1.5}Y_{0.5}Hf_{1.5}(PO4)3}&1&0.608&None&0.2, [0.1, 0.4]&$0.338\pm0.013$&SSE\\
 \ch{LiTa_{1.5}Nb_{0.5}Si2PO12}&20&0.613&\makecell{(Li$_1$-Li$_{1.5}$: 1.1)\\(Li$_{1.5}$-Li$_2$: 0.6)}&7e-5, [2e-5,3e-4]&$0.595\pm0.035$&SSE\\
 \ch{Li3ScIn(PO4)3}&14&0.613&None&6e-5, [1e-5, 4e-4]&$0.601\pm0.042$&SSE\\
 \ch{Li_{1.5}HfZrSi_{0.5}P_{2.5}O12}&13&0.620&None&1.2, [0.7, 1.8]&$0.282\pm0.011$&SSE\\
 \ch{Li_xHf_{1.5}W_{0.5}SiP2O12}&16 (Li$_{1.5}$)&0.775&\makecell{Li$_1$-Li$_{1.5}$: 3.1\\(Li$_{1.5}$-Li$_{2}$: 1.9)}&\makecell{0.7, [0.4, 1.1]\\(Li$_{1.5}$)}&\makecell{$0.301\pm0.011$\\(Li$_{1.5}$)}&MIEC\\
 \ch{Li_xHf_{1.5}Fe_{0.5}(PO4)3}&0 (Li$_{1.5}$)&0.846&\makecell{(Li$_1$-Li$_{1.5}$: 4.8)\\Li$_{1.5}$-Li$_{2}$: 2.5}&\makecell{7.2, [4.8, 10.8]\\(Li$_{1.5}$)}&\makecell{$0.217\pm0.010$\\(Li$_{1.5}$)}&MIEC\\
 \ch{Li_xHf_{1.5}Co_{0.5}(PO4)3}&9 (Li$_{1.5}$)&0.865&\makecell{(Li$_1$-Li$_{1.5}$: 4.7)\\Li$_{1.5}$-Li$_{2}$: 3.8}&\makecell{7.3, [4.9, 11.0]\\(Li$_{1.5}$)}&\makecell{$0.216\pm0.010$\\(Li$_{1.5}$)}&MIEC\\
 \ch{Li_xSc_{1.5}W_{0.5}(PO4)3}&23 (Li$_{2}$)&0.907&\makecell{Li$_{1.5}$-Li$_2$: 3.1\\Li$_{2}$-Li$_{2.5}$: 2.7}&\makecell{6.0, [3.9, 9.2]\\(Li$_{1.5}$)}&\makecell{$0.209\pm0.010$\\(Li$_{1.5}$)}&MIEC\\
 \ch{Li_xSc_{1.5}Mo_{0.5}(PO4)3}&13 (Li$_{3}$)&0.934&\makecell{Li$_{1.5}$-Li$_2$: 3.5\\Li$_{2}$-Li$_{2.5}$: 3.4}&\makecell{1.0, [0.6, 1.6]\\(Li$_{2}$ \footnotemark[2])}&\makecell{$0.302\pm0.011$\\(Li$_{2}$ \footnotemark[2])}&MIEC\\
 \ch{Li_xHf_{1.5}Mn_{0.5}(PO4)3}&0 (Li$_{1.5}$)&1.041&\makecell{(Li$_1$-Li$_{1.5}$: 4.3)\\Li$_{1.5}$-Li$_{2}$: 3.1}&\makecell{1.7, [1.1, 2.7]\\(Li$_{1.5}$)}&\makecell{$0.274\pm0.011$\\(Li$_{1.5}$)}&MIEC\\
 \ch{Li_xHf_{1.5}V_{0.5}Si_{0.5}P_{2.5}O12}&16 (Li$_{2}$)&1.063&\makecell{Li$_1$-Li$_{1.5}$: 3.7\\Li$_{1.5}$-Li$_{2}$: 3.1\\(Li$_{2}$-Li$_{2.5}$: 1.3)}&\makecell{5.3, [3.6, 7.9]\\(Li$_{1.5}$)}&\makecell{$0.231\pm0.010$\\(Li$_{1.5}$)}&MIEC\\
\end{tabular}
\end{ruledtabular}
}
\footnotetext[1]{The experimental conductivity data is obtained from \cite{VenkateswaraRao2015_LTP, Paris1996_LTP}}
\footnotetext[2]{The composition with x = 1.5 (\ch{Li_{1.5}Sc_{1.5}Mo_{0.5}(PO4)3}) has E$_{hull}$= 28 meV/atom, which exceeds the DFT-screening cutoff value (25 meV/atom), and thus is not selected for the MD simulations}
\label{tab:subset_candidates_NASICON} 
\end{table*}

\begin{table*}
\caption{\label{tab:table3} \textbf{A subset of the final screened garnet candidates} All identified garnet MIECs and 11 out of the 121 final garnet SSE candidates are tabulated here. For each MIEC compound, Li-ion conductivity is evaluated only at the composition specified in parenthesis. This composition satisfies all DFT-screening criteria and has a Li content x closest to 7.0, which has been experimentally shown to yield higher Li-ion conductivity \cite{Thangadurai2014_garnet_review} in garnet. For additional details, refer to Table \ref{tab:subset_candidates_NASICON}.}
{\fontsize{8}{10}\selectfont
\renewcommand{\arraystretch}{2.0}
\rowcolors{2}{gray!20}{white}
\begin{ruledtabular}
\begin{tabular}{ccc
>{\centering\arraybackslash}p{2.4cm}
>{\centering\arraybackslash}p{2.5cm}
>{\centering\arraybackslash}p{2.3cm}
c}
 Compound&\makecell{$E_{hull}$\\(meV/atom)}&\makecell{Max $\Delta\phi_{pbx}$\\(eV/non-Li/O/H)}
&\makecell{Li intercalation\\voltage (V)}&\makecell{Li-ion conductivity \\at 300 K (mS/cm)}&\makecell{Activation\\energy (eV)}&Category\\ \hline
\ch{Li7La3Zr2O12}&7&0.822&None&0.31 (Exp.)\footnotemark[1]&0.34 (Exp.)\footnotemark[1]&\makecell{Baseline\\SSE}\\
\ch{Li3La3W2O12}&0&0.282&\makecell{(Li$_3$-Li$_{5}$: 1.1)\\(Li$_{5}$-Li$_7$: 0.9)}&None&None&SSE\\
\ch{Li_{3.5}La3Ta_{0.5}W_{1.5}O12}&0&0.330&\makecell{(Li$_{3.5}$-Li$_{4.5}$: 1.2)\\(Li$_{4.5}$-Li$_{5}$: 1.1)\\(Li$_{5}$-Li$_{6.5}$: 0.9)\\(Li$_{6.5}$-Li$_7$: 0.8)}&1.3, [0.7, 2.4]&$0.233\pm0.015$&SSE\\
\ch{Li_{3.5}La3Nb_{0.5}W_{1.5}O12}&0&0.340&\makecell{(Li$_{3.5}$-Li$_{4.5}$: 1.2)\\(Li$_{4.5}$-Li$_{5}$: 1.1)\\(Li$_{5}$-Li$_{6.5}$: 0.9)\\(Li$_{6.5}$-Li$_7$: 0.6)}&1.7, [0.9, 3.3]&$0.222\pm0.016$&SSE\\
\ch{Li4La3TaWO12}&0&0.363&\makecell{(Li$_{4}$-Li$_{5}$: 1.1)\\(Li$_{5}$-Li$_{6}$: 0.8)}&0.2, [0.1, 0.4]&$0.323\pm0.016$&SSE\\
\ch{Li4La3NbWO12}&0&0.382&\makecell{(Li$_{4}$-Li$_{5}$: 1.1)\\(Li$_{5}$-Li$_{6}$: 0.72)\\(Li$_{6}$-Li$_{7}$: 0.70)}&0.3, [0.1, 0.5]&$0.319\pm0.014$&SSE\\
\ch{Li5La3Ta2O12}&8&0.397&None&3.9, [2.5, 6.3]&$0.228\pm0.011$&SSE\\
\ch{Li5La3Ta_{1.75}Nb_{0.25}O12}&9&0.408&None&1.9, [1.2, 3.2]&$0.260\pm0.012$&SSE\\
\ch{Li6La3Sc_{0.5}Ta_{1.5}O12}&23&0.624&None&2.5, [1.6, 4.1]&$0.246\pm0.012$&SSE\\
\ch{Li6La3ZrTaO12}&17&0.644&None&5.6, [3.5, 8.8]&$0.210\pm0.010$&SSE\\
\ch{Li_{6.5}La3Zr_{1.5}Ta_{0.5}O12}&12&0.728&None&10.7, [7.1, 16.4]&$0.188\pm0.010$&SSE\\
\ch{Li_{6.5}La3Hf_{1.5}Ta_{0.5}O12}&15&0.732&None&13.1, [8.5, 20.2]&$0.178\pm0.010$&SSE\\

\ch{Li_xLa3Cr_{0.25}W_{1.75}O12}&0 (Li$_{3.5}$)&0.464&
\makecell[c]{Li$_{3}$-Li$_{3.5}$: 2.8\\(Li$_{3.5}$-Li$_{4.5}$: 1.8)\\(Li$_{4.5}$-Li$_{5}$: 1.2)\\(Li$_{5}$-Li$_{5.5}$: 1.11)\\(Li$_{5.5}$-Li$_{6.5}$: 1.09)\\(Li$_{6.5}$-Li$_7$: 1.0)}&
\makecell{1.6, [0.9, 2.9]\\(Li$_{3.5}$)}&\makecell{$0.230\pm0.014$\\(Li$_{3.5}$)}&MIEC\\

\ch{Li_xLa3Fe_{0.25}W_{1.75}O12}&0 (Li$_{4}$)&0.530&\makecell{Li$_{3.5}$-Li$_{4.0}$: 2.0\\(Li$_{4}$-Li$_{4.5}$: 1.2)\\(Li$_{5.5}$-Li$_{7}$: 1.1)}&\makecell{0.3, [0.2, 0.7]\\(Li$_{4.0}$)}&\makecell{$0.284\pm0.017$\\(Li$_{4}$)}&MIEC\\
\ch{Li_xLa3Ni_{0.25}W_{1.75}O12}&1 (Li$_{4}$)&0.572&\makecell{Li$_{3.5}$-Li$_{4}$: 3.1\\(Li$_{4}$-Li$_{5.5}$: 1.2)\\(Li$_{5.5}$-Li$_{7}$: 1.1)}&\makecell{9e-3, [2e-3, 5e-2]\\(Li$_{4}$)}&\makecell{$0.402\pm0.040$\\(Li$_{4}$)}&MIEC\\
\ch{Li_xLa3Ta_{1.75}Bi_{0.25}O12}&7 (Li$_{5}$)&0.582&\makecell{Li$_{5}$-Li$_{5.5}$: 2.1}&\makecell{0.5, [0.3, 0.8]\\(Li$_{5.5}$)}&\makecell{$0.308\pm0.013$\\(Li$_{5.5}$)}&MIEC\\
\ch{Li_xLa3Co_{0.25}W_{1.75}O12}&0 (Li$_{3.5}$)&0.583&\makecell{Li$_{3.5}$-Li$_{4}$: 2.2\\(Li$_{4}$-Li$_{5.5}$: 1.2)\\(Li$_{5.5}$-Li$_{7}$: 1.1)}&\makecell{4e-2, [2e-2, 0.1]\\(Li$_{4}$)}&\makecell{$0.351\pm0.022$\\(Li$_{4}$)}&MIEC\\
\ch{Li_xLa3Cr_{0.5}W_{1.5}O12}&11 (Li$_{4}$)&0.591&\makecell{Li$_{3}$-Li$_{3.5}$: 3.0\\Li$_{3.5}$-Li$_{4}$: 2.7\\Li$_{4}$-Li$_{4.5}$: 2.4\\(Li$_{4.5}$-Li$_{6}$: 1.2)\\(Li$_{6}$-Li$_{6.5}$: 1.12)\\Li$_{6.5}$-Li$_{7}$: 1.10}&\makecell{7e-2, [3e-2, 0.2]\\(Li$_{4.5}$)}&\makecell{$0.359\pm0.022$\\(Li$_{4.5}$)}&MIEC\\
\ch{Li_xLa3Fe_{0.5}W_{1.5}O12}&8 (Li$_{5}$)&0.676&\makecell{Li$_{4}$-Li$_{4.5}$: 2.8\\(Li$_{4.5}$-Li$_{6}$: 1.2)\\(Li$_{6}$-Li$_{6.5}$: 1.13)\\(Li$_{6.5}$-Li$_{7}$: 1.06)}&\makecell{0.3, [0.1, 0.6]\\(Li$_{4.5}$)}&\makecell{$0.305\pm0.016$\\(Li$_{4.5}$)}&MIEC\\
\end{tabular}
\end{ruledtabular}
}
\footnotetext[1]{The experimental conductivity data is obtained from \cite{Li2012_garnet}}
\label{tab:subset_candidates_garnet} 
\end{table*}

The final NASICON and garnet compounds that pass all the screening filters in Figure \ref{fig:screening_scheme} are further categorized into possible SSE or MIEC cathodes based on their Li-insertion redox potential. More specifically, we assume that a compound that is redox-active within the battery operation voltage window ($2.0-4.2$ V vs. Li/\ch{Li+}) can potentially exhibit a reasonable electronic conductivity via a polaron hopping mechanism \cite{Franchini2021_polaron_review, Lin2025_MIEC_NASICON} between the mixed valence states present. This assumption is based on recent theoretical works that demonstrated generally low polaron hopping barriers ($<$ 0.45 eV) in garnets and NASICONs with mixed valence transition metals \cite{Schwarz2024_garnet_polaron, Lin2025_MIEC_NASICON}. The effectiveness of  substituting redox-active TMs to enhance electronic conductivity of an originally redox-inactive SSE has been previously experimentally validated. For instance, increases in electronic conductivity by 4 and 3 orders of magnitude were achieved for garnet-type \cite{Alexander2023_MIEC_garnet} and NASICON-type \cite{Scheiber2024_MIEC_LATP} Li-ion conductors, respectively. We estimate the redox activity by calculating the average Li intercalation voltages between the topotactically related compositions. Specifically, if there are any Li intercalation voltages that fall within the voltage range of $2.0-4.2$ V vs. Li/\ch{Li+}, the topotactically related compositions are grouped and categorized as a MIEC. Otherwise, the composition is considered to be redox-inactive and categorized as a SSE. In theory, the redox-active MIEC cathodes can provide extra Li storage capacity, though this is not our primary objective. 

We obtain 74 SSE and 7 MIEC candidates for NASICON, and 121 SSE and 7 MIEC candidates for garnet. All elements selected in these final candidates are presented in periodic tables in Figure \ref{fig:final_candidates}. The number in each box indicates the number of final candidate compounds containing the corresponding element. The corresponding substitution site is labeled by colors, where the color is scaled according to the number in the box, with darker colors indicating a more frequent appearance of the particular element in the final candidates. As can be seen, the elements in the final candidates are not uniformly present across the periodic table. For NASICON, the early TMs, including Sc, Ti, Zr, and Hf, are selected more frequently as the cations to occupy the octahedral M site, while Nb, Ta, and W are favored for garnet. Also, all final NASICON candidates (74 SSEs and 7 MIECs) contain phosphate groups, with some of them substituted with silicate groups. The complete list of compositions and their predicted stabilities of all final SSE candidates are provided in the Supplemental Table.

In Tables \ref{tab:subset_candidates_NASICON} and \ref{tab:subset_candidates_garnet}, we list a subset of the final SSE candidates (10 NASICON SSEs and 11 garnet SSEs) and all MIEC candidates (7 NASICON MIECs and 7 garnet MIECs), along with their DFT-predicted properties. The baseline compounds LTPO and LLZO are also included in the table for comparison. The subset SSEs are selected as follows: For NASICON, the ten most alkaline-stable compounds with the lowest values of max $\Delta\phi_{pbx}$ are selected. For garnet, we selected 11 SSEs, which include the three most alkaline-stable garnets with the lowest max $\Delta\phi_{pbx}$ values among all SSEs, the two most alkaline-stable garnets with 4 Li per formula unit (Li4-garnet), the two most alkaline-stable garnets with 5 Li per formula unit (Li5-garnet), the two most alkaline-stable garnets with 6 Li per formula unit (Li6-garnet), and the two most alkaline-stable garnets with 6.5 Li per formula unit (Li6.5-garnet). The high Li content garnet (Li content larger than 3.0) is intentionally included because the Li-stuffed garnets generally exhibit higher Li-ion conductivity \cite{Thangadurai2014_garnet_review, Xiao2021_NASICON_garnet}. For this subset of final candidate compounds, we additionally calculated the room temperature Li-ion conductivity with MD simulations using fine-tuned CHGNet models\cite{Jun2024_nitride,Lian2025_finetuned_CHGNet_NEB} (see Supplemental Information for fine-tuning methodologies) and the results are listed in the same Tables. The Li-ion conductivity result will be discussed more in detail in a later section.

The data in Table \ref{tab:subset_candidates_NASICON} shows that the maximum $\Delta\phi_{pbx}$ can be reduced up to 48\% for a pure Sc-based NASICON compound \ch{Li3Sc2(PO4)3} compared to the baseline value of LTPO. An even more significant improvement is achieved for the garnet compounds (Table \ref{tab:subset_candidates_garnet}), with the pure W-based garnet \ch{Li3La3W2O12} achieving the maximum decrease of 66\% in max $\Delta\phi_{pbx}$ relative to the baseline value of LLZO. In the following discussion section, we provide more in-depth analyses of the screening results, rationalizing the origin of high alkaline stability of specific chemistries. Potential trade-offs between alkaline stability and other material properties will also be discussed, offering design principles for optimizing compositions to achieve multiple desirable material properties.

\subsection{Effect of octahedral-site substitution on alkaline stability}
\label{subsec:oct_alk} 
\begin{figure*}[t]
\centering
\includegraphics[width=\linewidth]{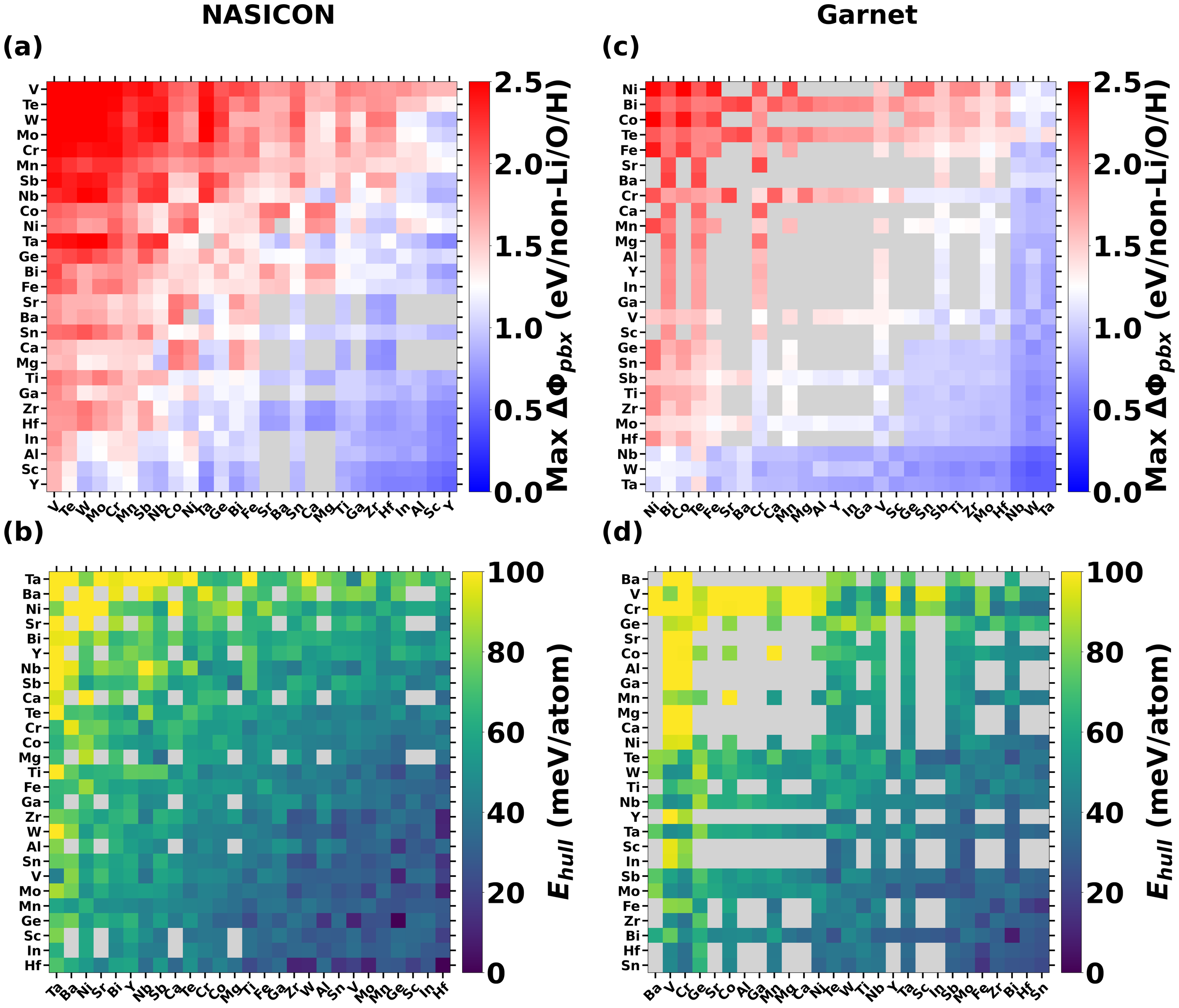}
\caption{\textbf{Effect of the type of cation-pair in the octahedral site on stability} The median values of max $\Delta\phi_{pbx}$ and $E_{hull}$ for different cation pairs in the octahedral sites of NASICON are shown in (a) and (b), respectively. Similarly, the corresponding median values for garnet are shown in (c) and (d), respectively. Elements are ordered along the axes based on the average values along each column or row. Only phosphate-based NASICONs are considered in (a) and (b). For both NASICON and garnet, the median values of max $\Delta\phi_{pbx}$ in (a) and (c) are calculated by including only compounds with $E_{hull} < 100$ meV/atom to exclude intrinsically unstable compounds. If no compound with $E_{hull} < 100$ meV/atom exists for a given cation combination, the box is greyed out. Grey boxes in (b) and (d) indicate that the corresponding cation combinations are not considered in our screening due to violations of the compositional constraints on the range of Li content or charge neutrality. All values are based on CHGNet-relaxed energies.}
\label{fig:oct_site_substitution} 
\end{figure*}

Charge neutrality requires that not every cation can be substituted into each composition. For this reason, we plot in Figure \ref{fig:oct_site_substitution} the average effect of the allowed cation pairs on the alkaline stability (max $\Delta\phi_{pbx}$) and synthesizability ($E_{hull}$). The cations are ordered along each axis based on the average values along each column or row. Therefore, in Figure \ref{fig:oct_site_substitution} (a) and (c), the overall alkaline stability of the substituted compounds increases from the upper left corner to the lower right corner. Similarly, the $E_{hull}$ increases, and thus the compound becomes less synthesizable from the upper left corner to the lower right corner in Figure \ref{fig:oct_site_substitution} (b) and (d).

For NASICON, the results shown in Figure \ref{fig:oct_site_substitution} (a) indicate that substituting early TMs, such as Y, Sc, Hf, Zr, and Ti, generally results in lower max $\Delta\phi_{pbx}$ values, and thus are beneficial for improving alkaline stability. In addition, all early TMs also have a high elemental PI of 1.0 (see Figure \ref{fig:stability_metric} (e)), indicating NASICON compounds substituted with early TMs can be potentially protected by the stable surface passivations in alkaline environment. These early TMs are therefore also predominant in the final candidate list, as depicted in Figure \ref{fig:final_candidates} (a). The exception is for Y-based NASICONs: while we obtained more than 18 final candidates for Sc-, Hf-, Zr-, and Ti-based NASICONs, as shown in Figure \ref{fig:final_candidates} (a), only 3 Y-based NASICONs passed the final screening. The lower pass rate for Y-based NASICON originates from its relatively high $E_{hull}$ value, which is evident from the lighter color shown in the column of Y in Figure \ref{fig:oct_site_substitution} (b). Our DFT calculation indicates that \ch{Li3Y2(PO4)3} lies 66 meV/atom above the convex hull, being less stable than  \ch{Li3Sc2(PO4)3}, \ch{LiHf2(PO4)3}, \ch{LiZr2(PO4)3}, and \ch{LiTi2(PO4)3} with $E_{hull}$ value of  14, 0, 18, and 0 meV/atom respectively. To the best of our knowledge, the synthesis of \ch{Li3Y2(PO4)3} has not been experimentally reported so far.

Figure \ref{fig:oct_site_substitution} (a) also shows that post-TM (Al, Ga, and In)-based NASICONs exhibit low max $\Delta\phi_{pbx}$, with Al- and In-based NASICONs ranking even higher than those based on Hf and Zr. However, these post-TMs are less common in final candidates compared to the compounds containing early TMs. This is due to the low elemental PI of these cations, as indicated by elemental PI values of 0.28, 0.0, and 0.04 for Al, Ga, and In, respectively in Figure \ref{fig:stability_metric} (e). Since phosphorus in the phosphate group does not provide any passivation capability with its elemental PI = 0.0, the NASICON compounds including post-TMs, such as Al, Ga, and In, must be paired with other cations with high elemental PI to have a stable surface passivation layer and to prevent continuous dissolution of the bulk material. Such chemistry-dependent passivation capability is the reason why we observe the broad distribution of PI for NASICON in Figure \ref{fig:chgnet_screening} (e).

Another interesting compositional category is the NASICONs based on alkaline earth metals (Mg, Ca, Sr, and Ba). Similar to the post-TMs, these alkaline earth metals also have low elemental PI values, with PI values of 0.0 for Mg, 0.0 for Ca, 0.09 for Sr, and 0.06 for Ba, as shown in Figure \ref{fig:stability_metric} (e). Therefore, alkaline earth metal are rare in the final NASICON candidate list, as shown in Figure \ref{fig:final_candidates} (a). However, as can be seen from Figure \ref{fig:oct_site_substitution} (a), the NASICON compounds that are substituted with a combination of an alkaline earth metal and an early transition metal, such as Mg-Hf, Mg-Zr, and Mg-Ti cation pairs, have lower driving force for decomposition in an alkaline solution, with a value of max $\Delta\phi_{pbx}$ even lower than the pure Hf-, Zr-, and Ti-based NASICONs. According to our DFT calculations, the values of max $\Delta\phi_{pbx}$ are 0.595, 0.642, and 0.703 eV/non-Li/O/H for the compounds \ch{Li3MgHf(PO4)3}, \ch{Li3MgZr(PO4)3}, and \ch{Li3MgTi(PO4)3}, respectively. In comparison, the values of max $\Delta\phi_{pbx}$ for the pure early TM-based NASICONs are relatively higher, with values of 0.697, 0.753, and 0.963 eV/non-Li/O/H for \ch{LiHf2(PO4)3}, \ch{LiZr2(PO4)3}, and \ch{LiTi2(PO4)3}, respectively. On the other hand, the NASICON compounds based on these alkaline earth metals generally have high $E_{hull}$ values, making these compounds likely more challenging to be synthesized. For instance, the DFT-estimated $E_{hull}$ values are 26, 46, 32 for \ch{Li3MgHf(PO4)3}, \ch{Li3MgZr(PO4)3}, and \ch{Li3MgTi(PO4)3}, respectively, all of which exceed the 25 meV/atom cutoff in the DFT-screening step. Indeed, it was experimentally reported that solid-state synthesis of \ch{Li_{1+2x}Mg_xZr_{2-x}(PO4)3} series results in the formation of impurty phases for x $>0.2$ \cite{Zhou2020_Mg-LZP}. This is another factor contributing to the scarcity of the alkaline earth metal-based NASICONs in the final candidate list. Nevertheless, the results presented above for both alkaline earth metal- and post-TM-based NASICONs indicate that while substituting low-PI elements might lead to a poorer passivation, it does not necessarily increase the driving force for decomposition in alkaline media (max $\Delta\phi_{pbx}$) and may even reduce it. The distributions of all types of stability metrics for each early TM, post-TM, and alkaline earth metal mentioned above are provided in Supplemental Information Figure S\ref{fig:oct_nasicon}.

Among the garnets, Figure \ref{fig:oct_site_substitution} (c) shows that W-, Ta-,and Nb-based compounds have significantly lower values of max $\Delta\phi_{pbx}$ than compounds based on other cations. Similar to their beneficial effects on alkaline stability in NASICON, occupation of octahedral sites by the early TMs, such as Hf, Zr, and Ti, also tends to enhance the alkaline stability in garnet. Additionally, Mo- and Sb-based garnets also demonstrate alkaline stability comparable to the compounds with early TMs. These results are consistent with the final screening results shown in Figure \ref{fig:final_candidates}(b), where W-, Ta-, and Nb-based garnets are the most common, followed by compounds containing early TMs (especially Hf, Zr, and Ti), as well as Mo and Sb. The distributions of all types of stability metrics for these elements that increases garnet's alkaline stability mentioned above are provided in Supplemental Information Figure S\ref{fig:oct_garnet}.

In contrast to NASICON compounds, where conductors with low driving forces for decomposition in alkaline solution ($\Delta\phi_{pbx}$) typically contain early transition metals with high elemental PI values, many of the most common octahedral dopants in garnets exhibiting low $\Delta\phi_{pbx}$ have low elemental PI values. For instance, while Sb has a relatively high PI of 0.7, W, Nb, and Mo show much lower values of 0.01, 0.0, and 0.09, respectively (Figure \ref{fig:stability_metric}e).
All of these values are considerably lower than the early TMs (elemental PI = 1.0), and are also lower than the cutoff values (PI $>$ 0.8) applied in both CHGNet pre-screening and DFT-screening steps. The high pass rate of these garnets substituted with elements having low elemental PI values can be attributed to the presence of La (elemental PI = 0.84) in the composition, which ensures that the compound PI remains high. This is evidenced by the narrow distributions of PI peaking near 1.0 across all La-based garnet compounds considered in our screening as shown in Figures \ref{fig:chgnet_screening} (h). 

\begin{figure*}[t]
\centering
\includegraphics[width=\linewidth]{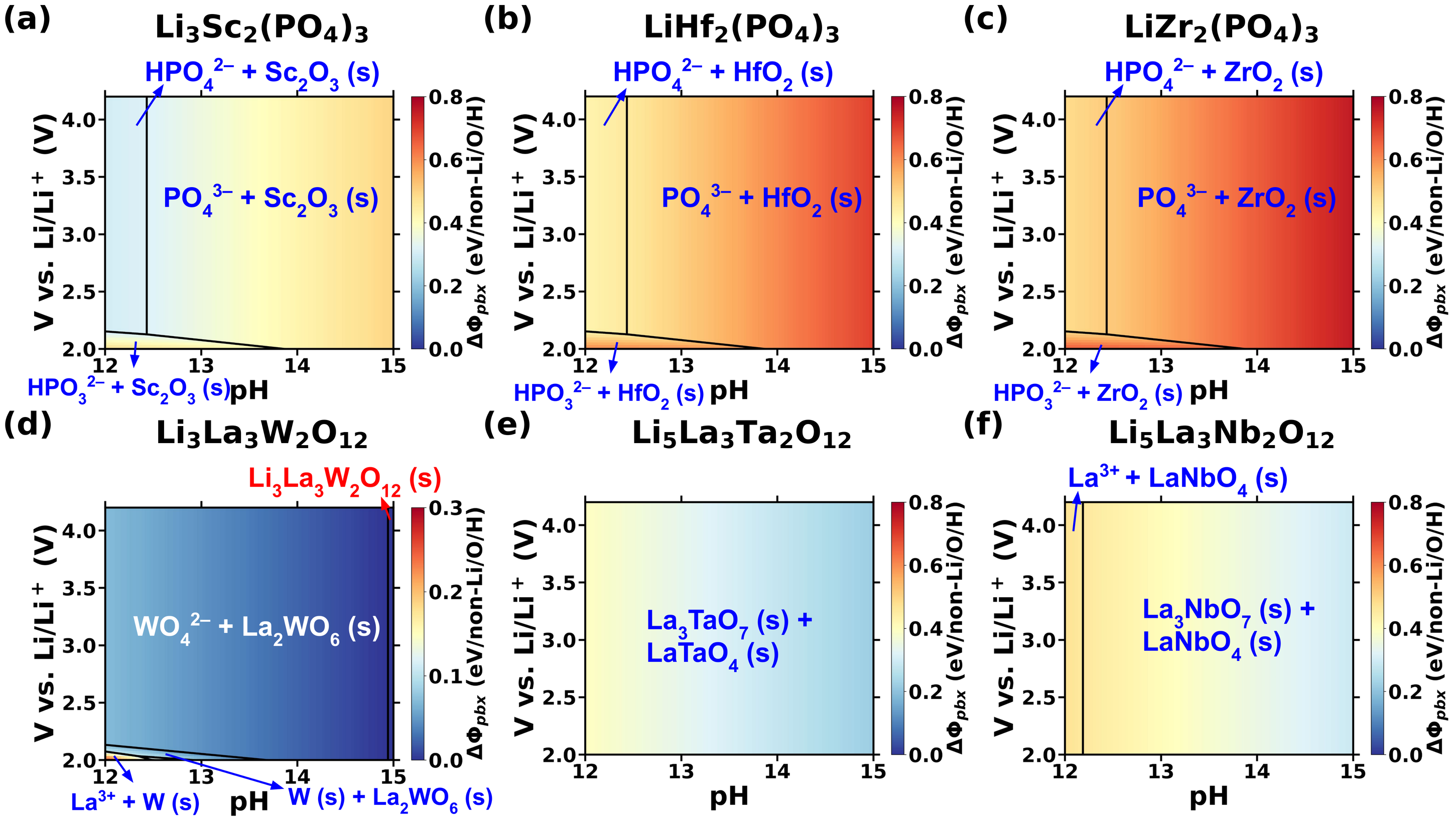}
\caption{\textbf{Pourbaix diagrams of alkaline stable Li-ion conductors} (a-c) NASICON, (d-f) garnet. The Pourbaix diagrams are constructed with DFT energies, following the correction scheme of Persson et. al. \cite{Persson2012_pbx}. The color scheme shows the values of $\Delta\phi_{pbx}$ as a function of pH and V. The most stable decomposition products at different pH and V conditions are also labeled. Note that the color scheme in (d) is scaled differently from other plots for clarity purpose. }
\label{fig:pbx_diagram} 
\end{figure*}

To better understand the underlying mechanisms for the low driving force for decomposition in alkaline solutions and the passivation mechanisms, we show in Figure \ref{fig:pbx_diagram} Pourbaix diagrams of some representative NASICON and garnet compounds substituted with cations identified as most beneficial to alkaline stability in Figure \ref{fig:oct_site_substitution}. The Pourbaix diagrams are obtained by using DFT-relaxed energies using the referencing approach developed by Persson et al \cite{Persson2012_pbx}. The color scheme in the plots shows the variations of $\Delta\phi_{pbx}$ as a function of pH and voltage. The most stable decomposition products (either solid or ionic solution species) at different pH and voltage conditions are labeled in each Pourbaix diagram.

We first discuss the results for NASICON compounds. As shown in Figures \ref{fig:pbx_diagram} (a-c), \ch{Li3Sc2(PO4)3} exhibits lower $\Delta\phi_{pbx}$ compared to \ch{LiHf2(PO4)3} and \ch{LiZr2(PO4)3} and is thus likely to be more stable. This agrees with the results shown in Figure \ref{fig:oct_site_substitution} (a) that Sc-based NASICON compounds in general have a lower max $\Delta\phi_{pbx}$ than Hf-based and Zr-based NASICONs. The Pourbaix diagram also suggests that all three NASICON compounds exhibit higher values of $\Delta\phi_{pbx}$ as pH increases. 

The decomposition products in the Pourbaix diagrams show that the dissolution of the \ch{PO_4^{3-}} group drives the instability, leaving behind a solid TM oxide, mostly a binary oxide, at high pH conditions. The stability of the oxide decomposition is the reason why early TMs have the highest elemental PI of 1.0 (Figure \ref{fig:screening_scheme} (e)). It was recently experimentally demonstrated that the dissolution of phosphate group is indeed the major cause of the degradation of phosphate-based NASICON compounds in alkaline solutions \cite{Lam2024_LTGP_high_pH,Mishra2025_LATP}. We expect similar behavior from other types of polyanion groups, as the center cations in the polyanion groups, i.e., Si, S, V, and Mo, consistently exhibit low elemental PI values below 0.1, indicating their natural tendency to dissolve as ionic solution species at high pH conditions. Since none of these NASICON compounds are strictly stable, as indicated by the positive values of $\Delta\phi_{pbx}$, the bulk material can only be stabilized by surface passivation enabled by well-selected cation dopants in the octahedral sites.

In contrast, all garnet compounds considered in our screening contain lanthanum, which has a high elemental PI value of 0.84 (\ref{fig:stability_metric} (e)), implying that all of these La-based garnet compounds should potentially form La-oxide-based surface passivation upon decomposition, irrespective of the cations occupying the octahedral site. To examine this hypothesis, we constructed Pourbaix diagrams for W-, Ta-, and Nb-based garnets (\ch{Li3La3W2O12}, \ch{Li5La3Ta2O12}, and \ch{Li5La3Nb2O12}) as shown in Figures \ref{fig:pbx_diagram} (d-f). The hypothesis of La-based passivation is found to be mostly true, in a sense that all garnet compounds indeed decompose into La-based solid products for most of the pH and voltage ranges. However, instead of forming pure La-based binary oxides or hydroxides, all garnet compounds appear to decompose into La-M-O ternary oxides, where M is either W, Ta, or Nb. For example, the predominant solid products for \ch{Li3La3W2O12} is \ch{La2WO6}, while it is the combination of \ch{LaTaO4} and \ch{La3TaO7} and the combination of \ch{LaNbO4} and \ch{La3NbO7} for \ch{Li5La3Ta2O12} and \ch{Li5La3Nb2O12}, respectively. Similar results are obtained for Mo- and Sb-based garnets, which are predicted to form \ch{La2MoO6}, and the combination of \ch{LaSbO4} and \ch{La3SbO7}, respectively, as the predominant solid products at high pH conditions (see Supplemental Information Figure S\ref{fig:pbx_garnet_Mo_Sb}). Furthermore, Figure \ref{fig:pbx_diagram} (d) also indicates that at the highest considered pH conditions (pH$\approx$15), the garnet compound \ch{Li3La3W2O12} is predicted to remain intact without any decomposition. Our results indicate that an effective solid passivation is easier to be achieved for garnets than for NASICONs.

The substitutional metals (i.e., W, Ta, Nb, Mo, and Sb) that enhance garnet stability under alkaline conditions tend to have higher oxidation states (5+ or 6+) compared to \ch{Zr^{4+}} in the baseline compound \ch{Li7La3Zr2O12}. Consequently, garnets exhibiting a lower driving force to decompose in alkaline solutions ($\Delta\phi_{pbx}$) tend to have a lower Li content per formula unit (3 or 5 Li per 12 O). This raises the question of whether the reduced decomposition driving force originates primarily from the cation chemistry or from the lower Li content. As shown later in Figure \ref{fig:Li_conductivity} (c), the $\Delta\phi_{pbx}$ value indeed increases with Li content in garnet compounds. Although these two factors cannot be completely decoupled, our calculation indicate that low Li-content alone does not guarantee a low $\Delta\phi_{pbx}$ value. For example, CHGNet-predicted maximum $\Delta\phi_{pbx}$ values for \ch{Li3La3Te2O12}, \ch{Li3La3Cr2O12} and \ch{Li5La3V2O12} garnet compounds are 1.84, 1.41 and 1.13 eV/non-Li/O/H, respectively, all exceeding the screening cutoff value (see Table S\ref{tab:screen_criteria_garnet} in the Supplemental Information). Moreover, Figure \ref{fig:oct_site_substitution} shows that Te-, Cr-, V-substituted garnets are generally less stable under alkaline conditions. Overall, these results suggest that cation chemistry plays a more dominant role than Li content in governing the alkaline stability of garnet compounds.

The driving force for reaction in the highly alkaline environment ($\Delta\phi_{pbx}$) is substantially lower for garnets than for NASICONs. It can also be observed that $\Delta\phi_{pbx}$ of these garnets even slightly decrease as pH increases, exhibiting an opposite trend to that of the NASICON compounds. The variation of $\Delta\phi_{pbx}$ with pH can be rationalized by considering the change in the number of oxygen per chemical formula unit during decomposition. The grand Pourbaix potential $\phi_{pbx}$ can be mathematically expressed as (see Section S\ref{sec:methods_SI} in the Supplemental Information for the derivation)
\begin{align}
    \phi_{pbx} &= G - N_O\cdot\mu_{H_2O} - \left(2N_O - N_H\right)\cdot RT\cdot \ln10\cdot \text{pH} \notag\\
    &\quad -N_{Li}\cdot\mu_{Li^+}-\left[\left(2N_O-N_H-N_{Li}\right)\cdot e+Q\right]\cdot V
\label{equation_1}
\end{align}
\noindent where $G$ is the Gibbs free energy of a given phase, $N_O$, $N_H$, and $N_{Li}$ are the number of oxygen, hydrogen, and lithium in the compound, respectively, $\mu_{H_2O}$ is the chemical potential of a water molecule, and $\mu_{Li^+}$ is the Gibbs free energy of \ch{Li+} ion in the water, R is the gas constant, T is the temperature, e is the elemental charge, and Q is the charge for the ionic species. Since the grand Pourbaix potential $\phi_{pbx}$ varies linearly with $-N_O\cdot\text{pH}$, the grand Pourbaix decomposition energy $\Delta\phi_{pbx}$ also changes linearly with $-(N_{O,reactant} - N_{O,product})\cdot\text{pH} = -\Delta N_O\cdot\text{pH}$, where the subscript "reactant" denotes either NASICON or garnet compound, the subscript "product" refers to the most stable decomposition products at a given pH and voltage condition. Because the decomposition of NASICONs results in the formation of a soluble phosphate group (\ch{PO4^{3-}}) and a solid TM oxide (\ch{Sc2O3}, \ch{HfO2}, or \ch{ZrO2}) at pH $>$ 12.5 and most voltage values, the decomposition reaction absorbs additional oxygen from the environment. Take \ch{Li3Sc2(PO4)3} as an example (Figure \ref{fig:pbx_diagram} (a)), the decomposition reaction is \ch{Li3Sc2(PO4)3} $\rightarrow$ 3\ch{Li+} $+$ \ch{Sc2O3} $+$ 3\ch{PO4^{3-}}, and thus $\Delta N_O=-3$ per formula unit of \ch{Li3Sc2(PO4)3}. The negative sign of $\Delta N_O$ results in the $\Delta\phi_{pbx}$ values becoming more positive and the NASICON compounds being less stable as pH increases. In contrast, the primary decomposition products of garnets under most pH and voltage conditions are La-M-O ternary oxides (M = W, Ta, or Nb). The formation of ternary oxides results in the extraction of all lithium and the release of redundant oxygen upon decomposition, leading to positive values of $\Delta N_O$. It should be noted that the formation of \ch{WO4^{2-}} in the decomposition of \ch{Li3La3W2O12} actually consumes additional oxygen, similar to the formation of TM binary oxides for NASICONs. However, in this case, $\Delta N_O=1$ per formula unit of \ch{Li3La3W2O12} due to the following decomposition reaction: \ch{Li3La3W2O12} $\rightarrow$ 3\ch{Li^+} $+0.5$\ch{WO4^{2-}} $+1.5$\ch{La2WO6}. Consequently, $\Delta N_O$ remains positive for all of these garnet compounds, making them more stable as pH increases. Such variation in $\Delta\phi_{pbx}$ with pH explains the trend observed across the majority of pH and voltage ranges depicted in Figure \ref{fig:pbx_diagram}.

In summary, our results suggest that garnets not only have higher passivation capability, but also have lower driving force for dissolution in an alkaline solution than NASICONs, showing their superior potential as Li-ion conductors in high alkalinity.

\subsection{Effect of polyanion group substitutions in NASICONs on alkaline stability}

\begin{figure*}[t]
\centering
\includegraphics[width=\linewidth]{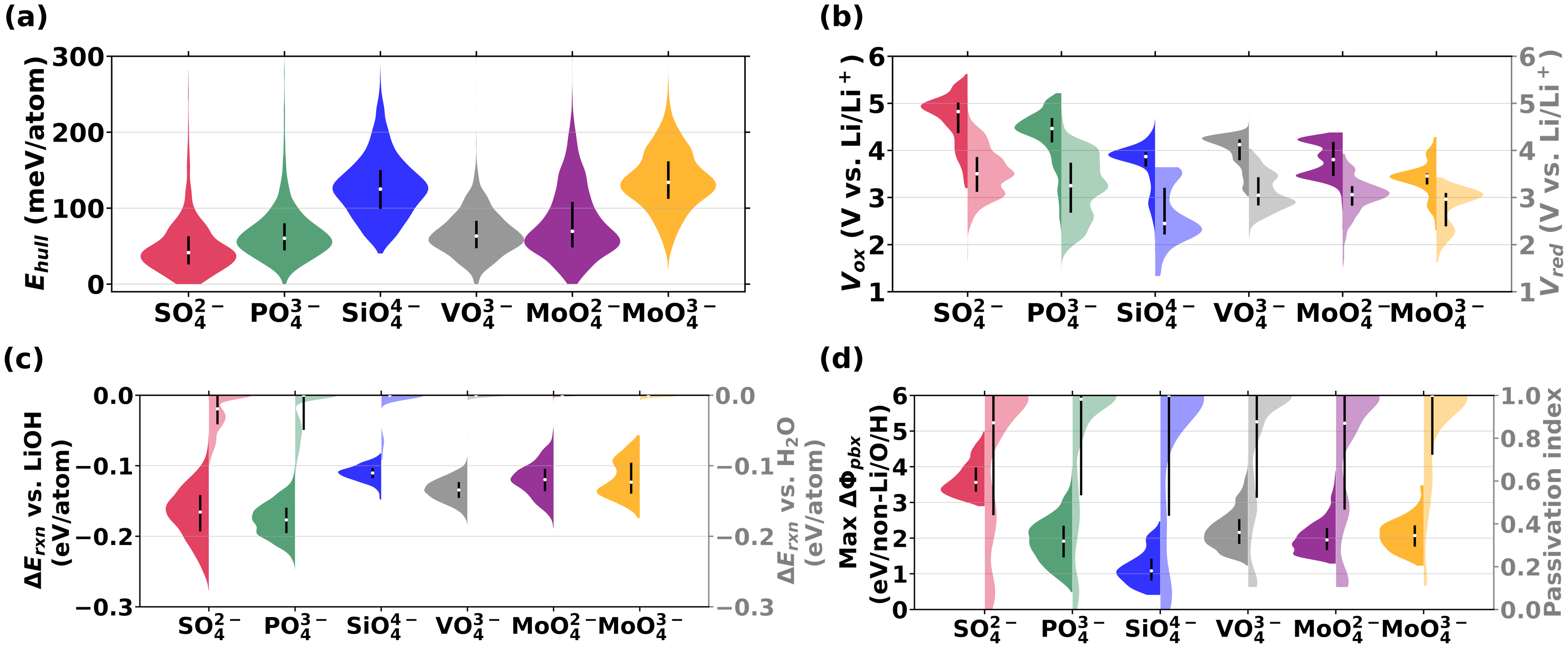}
\caption{\textbf{Effect of polyanion group on the stability of NASICON} (a) $E_{hull}$, (b) electrochemical stability, (c) chemical stability, and (d) alkaline stability. All stability metrics are based on CHGNet-relaxed energies. No compounds with mixed-polyanion groups are included. For the stability metrics other than $E_{hull}$, we only selected compounds with CHGNet-relaxed $E_{hull}$ $<$ 100 meV/atom to evaluate the trend within a set of reasonably stable compounds.}
\label{fig:polyanion_substitution} 
\end{figure*}

As can be seen from Figure \ref{fig:final_candidates} and Table \ref{tab:subset_candidates_NASICON}, there are many silicophosphate NASICONs in the final candidate list. To better understand how the stability of NASICON depends on the type of polyanion group, in Figure \ref{fig:polyanion_substitution} we show the distributions of stability metrics for NASICON compounds that contain different polyanion groups. Compounds with mixed polyanion groups are excluded to allow for a direct comparison of individual polyanion types.

Among the nonmetal cation-based polyanion groups, i.e., silicates (\ch{(SiO4)^{4-}}), phosphates (\ch{(PO4)^{3-}}), and sulfates (\ch{(SO4)^{2-}}), both $V_{ox}$ and $V_{red}$ decrease in the order of \ch{(SO4)^{2-}} $>$ \ch{(PO4)^{3-}} $>$ (\ch{(SiO4)^{4-}}, a trend previously attributed to the inductive effect \cite{Masquelier2013_polyanion_review}. The NASICONs with \ch{(SiO4)^{4-}} groups tend to exhibit lower $\Delta E_{rxn}$ against the reaction with LiOH compared to \ch{(SO4)^{2-}} and \ch{(PO4)^{3-}}. In terms of the alkaline stability, there is a clear decreasing trend in max $\Delta\phi_{pbx}$ following the order of \ch{(SO4)^{2-}} $>$ \ch{(PO4)^{3-}} $>$ (\ch{(SiO4)^{4-}}, making silicates the most stable under high pH conditions. However, the pure silicate-based NASICON compounds may be more challenging to synthesize due to their higher $E_{hull}$ compared to other compounds with different polyanions. This is also reflected in the final candidates listed in Table \ref{tab:subset_candidates_NASICON} and in the Supplemental Table. Although there are many silicophosphtates, none of the pure silicate-based NASICON compounds passed the final stage of screening. 

For the polyanion groups with a TM center, both \ch{(VO4)^{3-}} and \ch{(MoO4)^{2-}} offer similar electrochemical and chemical stabilities as silicates, while exhibiting even lower $E_{hull}$. \ch{(MoO4)^{3-}} compounds also demonstrate comparable electrochemical and chemical stabilities to \ch{(VO4)^{3-}} and \ch{(MoO4)^{2-}}, but may be less synthesizable due to their higher $E_{hull}$. That said, all of these vanadates and molybdates have higher max $\Delta\phi_{pbx}$ values than the silicates and some of the best phosphates. Our screening prioritize the compounds with higher alkaline stability. As a result, no molybdate-based and only few vanadate-substituted NASICONs remained in the final candidate list.

\subsection{Li conductivity vs. alkaline stability}

\begin{figure*}[t]
\centering
\includegraphics[width=0.6\linewidth]{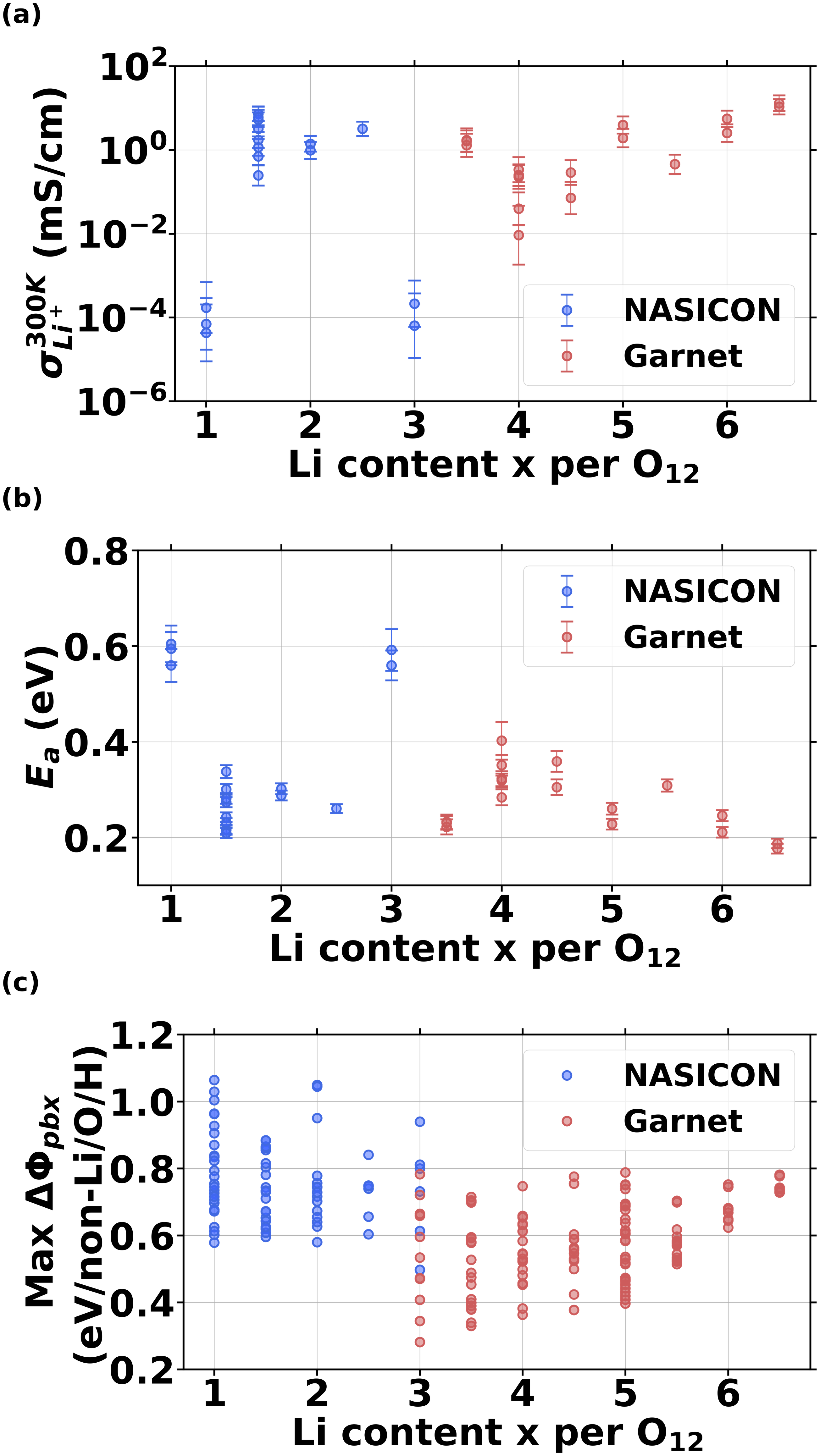}
\caption{\textbf{Effect of Li content on the Nernst-Einstein Li-ion conductivity and alkaline stability} (a) Li-ion conductivity at 300 K ($\sigma_{\ch{Li+}}^{300 K}$) is extrapolated from the high-temperature conductivity obtained by fine-tuned CHGNet MD simulations at high temperature (500-1000 K).  (b) Activation energy ($E_a$) of Li+ self diffusion coefficient, (c) max $\Delta\phi_{pbx}$ of all final SSE candidates (74 NASICON SSEs and 121 garnet SSEs) are plotted as a function of Li content. MIECs are excluded from the plots because the Li content of a MIEC varies as a function of applied voltage. The values of max $\Delta\phi_{pbx}$ are obtained from DFT-relaxed energies}
\label{fig:Li_conductivity} 
\end{figure*}

For the compounds listed in Tables \ref{tab:subset_candidates_NASICON} and \ref{tab:subset_candidates_garnet}, the room-temperature Li-ion conductivities are evaluated by performing MD simulations using the fine-tuned CHGNet model (see the Supplemental Information for the details on fine-tuning). More specifically, we run 1.4 ns-long MD simulations at high temperatures (500-1000 K), from which the Nernst-Einstein Li-ion conductivity at 300 K is extrapolated by assuming an Arrhenius relationship. For each MIEC compound, Li-ion conductivity is evaluated only for the composition that meets all DFT-screening criteria and has a Li content (x) closest to 1.5 for NASICON compounds and 7.0 for garnet compounds. These Li contents were experimentally shown to result in higher Li-ion conductivities \cite{Rossbach2018_NASICON_review, Thangadurai2014_garnet_review}. The CHGNet-predicted 300 K Li-ion conductivity and activation energy are plotted as a function of Li content in Figures \ref{fig:Li_conductivity} (a) and (b). 

Figure \ref{fig:Li_conductivity} (a) shows that, consistent with experiments \cite{Rossbach2018_NASICON_review}, NASICON compounds with Li content x = 1.5 achieve the highest room-temperature Li-ion conductivity. The high Li-ion conductivity is maintained for x = 2.0 or 2.5, whereas the compounds at two end-point compositions with x = 1.0 or 3.0 have much lower conductivity. The activation energy shows a similar trend, with $E_a$ below 0.35 eV at 1.5 $\le$ x $\le$ 2.5, while activation energies exceed 0.5 eV for x = 1.0 or 3.0 (see Figure \ref{fig:Li_conductivity} (b)). For garnet compounds, our simulation results agree with the experimental trend that the Li-ion conductivity generally increases with increasing Li content \cite{Thangadurai2014_garnet_review}.

To investigate if both high Li-ion conductivity and high alkaline stability can be achieved in the same composition, we plot in Figure \ref{fig:Li_conductivity} (c) max $\Delta\phi_{pbx}$ of all final SSE candidates (74 NASICON SSEs and 121 garnet SSEs), including the compounds considered in Figures \ref{fig:Li_conductivity} (a) and (b), as a function of Li content. The results show different trends between NASICON and garnet compounds: the NASICON compounds do not exhibit a clear dependence of max $\Delta\phi_{pbx}$ on the Li content, while the lowest max $\Delta\phi_{pbx}$ value increases with Li content for the garnet compounds. Our results suggest that for NASICON compounds, the Li-ion conductivity can potentially be optimized by tuning Li content without sacrificing the alkaline stability. However, for garnet compounds, Li stuffing increases Li-ion conductivity at the expense of the alkaline stability, posing a fundamental challenge for designing garnet-type conductors with well-balanced alkaline stability and ionic conductivity.


\subsection{Electronic conductivity vs. alkaline stability}

The results shown in Figure \ref{fig:oct_site_substitution} reveals that the first-row TMs, such as V, Cr, Mn, Fe, Co, and Ni, generally lead to high max $\Delta\phi_{pbx}$ values in both NASICON and garnet compounds, indicating that redox-active cations may destabilize the Li-ion conductors at high pH conditions. In Figure S\ref{fig:first-row-TM_substitution} in the Supplemental Information, we show the correlation between the amount of mixed-valent TMs in the NASICON or garnet compound and max $\Delta\phi_{pbx}$ values. The result indeed shows that max $\Delta\phi_{pbx}$ value increases by a factor of 2-3 as more mixed-valent TM cations are substituted (y decreases). Limiting the amount of TM may be a strategy to stabilize NASICON or garnet compounds, but to achieve good electronic conductivity, a high enough concentration of redox-active metals along which polarons can hop has to be maintained \cite{Lee2024_Mn_polaron_percolation, Lin2025_MIEC_NASICON}. This suggests that achieving both high electronic conductivity and high alkaline stability may be intrinsically challenging. As we prioritize the compounds with higher alkaline stability in our screening, we obtained much more SSE (195 SSEs) compared to MIECs (14 MIECs) in the final candidate list. Also, the selected MIEC compounds have relatively low concentrations of the redox-active TMs.

\section{Discussion}

Oxide-based solid-state Li-ion conductors, especially NASICON and garnet-type conductors, have been widely studied for decades, and their crystal structures \cite{Arbi2013_LATP,Catti2004_LFTP,Morgan2003_NASICON_structure,Delmas1988_NASICON, Callaghan2008_Li3, Callaghan2008_Li7}, Li-ion conduction mechanism \cite{Xiao2021_NASICON_garnet,Rossbach2018_NASICON_review,Thangadurai2014_garnet_review}, as well as electrochemical properties \cite{Richards2016_interface_stability} are relatively well understood. However, practical applications of these materials in batteries under operating conditions requires a combination of properties that is highly challenging to satisfy simultaneously. The goal of this work is to investigate the extent to which multiple key requirements for solid-state conductors can be optimized simultaneously. In particular, we focused on the trade-off between synthesizability, electrochemical stability, ionic conductivity, and stability in humid and alkaline environment. While these factors are broadly relevant to solid electrolytes, they are especially critical for solid-state humid Li-\ch{O2} batteries. By leveraging the pretrained universal machine-learning interatomic potentials, we were able to cover a vast chemical space and understand the broad trade-offs between properties in them. We successfully identified multiple NASICON and garnet compounds with properties that are superior to than their prototype compounds \ch{LiTi2(PO4)3} and \ch{Li7La3Zr2O12} (see Figure \ref{fig:chgnet_screening}). Moreover, our analysis points out the distinct advantages and disadvantages for each type of crystal framework, as summarized below.

First of all, the two material classes exhibit opposing stability profiles against \ch{H2O} and LiOH. NASICON is generally stable against water, which can remain intact after 8 months of immersion in distilled water \cite{Hasegawa2009_NASICON_water}. However, NASICON is vulnerable to LiOH or highly alkaline environments due to the dissolution of its polyanion groups in alkaline solutions \cite{Lam2024_LTGP_high_pH, Mishra2025_LATP}. By contrast, garnet is more resistant to alkaline decomposition due to the absence of polyanion phosphate groups, but it is highly sensitive to water. While not explicitly considered in our calculations, garnets can react with \ch{H2O}molecules through topotactic \ch{Li+}/\ch{H+} exchange \cite{Ye2021_garnet_LHX_review,Cheng2018_LHX,Sharafi2017_LHX_conductivity, Wang2024_garnet_LHX}. Our recent publication \cite{Li2026_LHX_oxide} shows that the proton exchange gets severer for Li-stuffed garnets with a higher Li content due to increased Li chemical potential. The driving force is so strong that the proton exchange can occur even in a saturated LiOH solution (pH=15) with quite low proton chemical potential.

Second, neither NASICON nor garnet compounds are strictly thermodynamically stable in alkaline environments, as both exhibit non-zero decomposition driving forces (positive max $\Delta\phi_{pbx}$ values). This suggests that their long-term stability might rely on the formation of a surface passivation layer. Accordingly, our screening targeted materials that combine a low decomposition driving force (low $\Delta\phi_{pbx}$) with a high tendency for surface passivation (high PI). For NASICON, the primary challenge is the dissolution of polyanion groups. Substituting silicate groups for phosphate groups can reduce the decomposition driving force in alkaline solutions, although it compromises synthesizability by increasing the $E_{hull}$. We also found that early transition metals, such as  Sc, Hf, Zr, and Ti, are the most effective dopants for NASICON materials, as they simultaneously lower the decomposition energy and promote surface passivation by forming stable oxides upon decomposition. In comparison, garnets benefit from the high PI values of their constituent lanthanide elements (see Figure \ref{fig:stability_metric} (e)). This relaxes the requirement for the octahedral cations to possess high PI values and allows compositional optimization toward lower  $\Delta\phi_{pbx}$ without affecting the passivation. As a result, we identified W, Ta, and Nb as the most beneficial elements, even though W and Nb are predicted to be low-PI elements. Therefore, we conclude that the superior alkaline stability of garnet compared to NASICON stems from two factors: a reduced thermodynamic driving force for decomposition due to the absence of polyanion groups, and the formation of lanthanide-based surface passivation layers.

Third, garnet faces a unique trade-off between high alkaline stability and high Li-ion conductivity that is absent in NASICON. Our result shows that the most stable garnet in an alkaline environment has the lowest Li content of 3 Li per formula unit, which is due to the incorporation of high oxidation state cations (e.g., \ch{W^{6+}}, \ch{Ta^{5+}}, and \ch{Nb^{5+}}). However, the low Li content also leads to negligible Li-ion conductivity at room temperature \cite{Xiao2021_NASICON_garnet}. While increasing the Li content to 6–7 per formula unit enables much higher conductivity, it also raises the driving force for alkaline decomposition by 2–3 times (see Table \ref{tab:subset_candidates_garnet} and Figure \ref{fig:Li_conductivity} (c)). A similar trade-off applies to the proton exchange reaction, where the high Li chemical potential due to Li stuffing promotes undesirable Li extraction into the solution \cite{Li2026_LHX_oxide}. Because these trade-offs are intrinsic to the garnet structure, designing a well-balanced garnet-type conductor likely requires strategies beyond simple compositional tuning.

In addition to the above-mentioned differences, both NASICON and garnet compounds face the same trade-off when trying to achieve high stability in alkaline conditions while retaining reasonable electronic conductivity, which may be particularly problematic when developing mixed conductors exposed to alkaline environments. This is because electronic conductivity requires some amount of redox-active cations, which are prone to dissolve in alkaline solutions. The exact amount of redox-active TM substitution required to achieve satisfactory electronic conductivity cannot be easily estimated, as direct evaluation of electronic conductivity (such as the evaluation of polaron formation energy and migration energy using \textit{ab-initio} methods \cite{Schwarz2024_garnet_polaron}) is not conducted in our screening procedure and does not always provide quantitatively accurate information.  A previous experimental study indicated that an increased electronic conductivity is achieved for MIEC garnets substituted with mixed-valent transition metals, with the electronic conductivity ranging from $10^{-5}$ to $10^{-6}$ S/cm \cite{Alexander2023_MIEC_garnet}. In comparison, the electronic conductivity was only increased to $10^{-9}$ S/cm by lithiating initially electronically insulating \ch{Li_{1.3}Al_{0.3}Ti_{1.7}(PO4)3} \cite{Scheiber2024_MIEC_LATP}, which is still orders of magnitude lower than the Li-ion conductivity. The low electronic conductivity of NASICON compounds is believed to be intrinsic in nature due to the long polaron hopping distance originating from the vertex-sharing polyanion tetrahedra and metal octahedra which limit direct hopping between equivalent sites \cite{Masquelier2013_polyanion_review}. One of the highest experimentally reported electronic conductivity of $9 \times 10^{-5}$ S/cm for NASICON is achieved by substituting V into the tetrahedral site of NASICON \cite{Sharma2021_V-NASICON}. However, V-substitution in tetrahedral sites of a phosphate-based NASICON is found to reduce its alkaline stability in our screening (see Figure \ref{fig:polyanion_substitution} (d)). As a result, only a few V-substituted NASICON compounds eventually passed the screening (see Figure \ref{fig:final_candidates} (a)). A more recent theoretical study pointed at Fe-substitution as effective in decreasing the polaron hopping barriers in lithium NASICON \cite{Lin2025_MIEC_NASICON}. The predicted elemental PI of Fe is quite high (0.81, see Figure \ref{fig:stability_metric} (e)), indicating that Fe-substitution may improve alkaline stability. We indeed obtained Fe-containing NASICON and garnet MIEC compounds (\ch{Li_xHf_{1.5}Fe_{0.5}(PO4)3} and \ch{Li_xLa3Fe_{0.25}W_{1.75}O12}) in the final candidate list. As experimental measurements of electronic conductivities for redox-capable NASICONs are still rare, we call for more theoretical and experimental exploration to examine whether chemical substitutions alone is enough to achieve electronic conductivity comparable to its ionic counterpart.

In summary, we conducted a hierarchical high-throughput screening of possible alkaline-stable Li-ion conductors combining both machine-learning interatomic potentials (CHGNet model) and DFT calculations. The pretrained CHGNet model enables pre-screening a vast range of chemistries, making it possible to chart almost the whole periodic table. Our analysis revealed that neither NASICON nor garnet are thermodynamically stable in alkaline environments, suggesting that surface passivation is essential for long-term stability. NASICON is subject to continuous degradation driven by phosphate group dissolution. In contrast, garnets exhibit a lower decomposition energy due to the absence of polyanion groups and the bulk material can be further stabilized by lanthanide-based oxide passivation layers. For each type of oxide conductors, we separately identified specific chemical substitutions that enhance stability by mitigating decomposition energies and promoting surface passivation. Our findings reveal fundamental trade-offs between material stability and conductivity, emphasizing the importance of multi-objective optimization for the practical deployment of oxide-based conductors, particularly in next-generation solid-state humid Li-\ch{O2} batteries.

\section{Acknowledgment}
This work was supported by the Samsung Advanced Institute of Technology (SAIT). This research used computational resources of the National Renewable Energy Laboratory (NREL) clusters under the \textit{saepssic} allocation. 
K.J. acknowledges support from Kwanjeong Educational Foundation Scholarship.

\section{Author Contributions}
Z.L and K.J. contributed equally to this work. Z.L and K.J. designed the screening procedures, and conducted all calculations. B.D. contributed to the fine-tuning of CHGNet models. The manuscript was written by Z.L. and K.J., and revised by B.D. and G.C. The work was supervised by G.C.

\section{Methods}
\label{sec:method_section}
\subsection{Substitution method}
For the NASICON framework (\ch{Li_xM2(AO4)3}), we used a hexagonal supercell with 72 O atoms, which has 12 octahedral sites in total. Up to 2 elements are allowed to substitute on octahedral sites, where the allowed substitution ratio include, 1/6, 1/4, 1/3, or 1/2. For garnet (\ch{Li_xLa3M2O12}), we used the supercell size with 96 O atoms, which has 16 octahedral sites in total. Similar to NASICON, up to 2 elements are allowed to substitute on octahedral sites, where the allowed substitution ratio include, 1/8, 1/4, 3/8, or 1/2. For all NASICON and garnet compounds with different Li contents and cation substitutions, we first ordered the Li/vacancy sites (6b. 36f, or 18e sites for NASICON; 24d or 96h for garnet), and then the cation substitution sites (M and A sites for NASICON; and M site only for garnet). The fractional occupation of Li ions in different Wycoff sites of NASICON are determined using two occupancy rules. The first rule involved populating lithium in 6b and 18e sites \cite{Delmas1988_NASICON}: across all Li content range ($1 \leq x \leq 3$), 6b sites are kept fully occupied, and the rest of Li occupies 18e sites. In the second rule, Li sequentially occupies 6b, 36f, and 18e sites as follows \cite{Arbi2013_LATP,Catti2004_LFTP,Morgan2003_NASICON_structure}: (1) For $x = 1$, all Li ions fully occupy 6b site. (2) For $1 < x \leq 2$, Li in 6b sites are gradually replaced by Li in 36f sites, such that Li in 6b sites become vacant at $x=2$. (3) For $2< x \leq 3$, the fractional occupation of 36f is kept at 1/3, and the rest of Li occupies 18e sites. For each Li content and occupancy rule, we obtained 20 lowest electrostatic Ewald energy Li/vacancy configurations with a fixed cation ordering using \texttt{pymatgen} \cite{Ong2013_pymatgen}. All the Li/vacancy configurations were relaxed using DFT calculations, from which the configuration with the lowest DFT-relaxed energy are selected. The fractional occupation of Li ions in different Wycoff sites of garnet are determined as follows \cite{Callaghan2008_Li3, Callaghan2008_Li7}: (1) For $x = 3$, all Li ions fully occupy the 24d site. (2) For $x = 7$, half of the 24d sites are occupied, and the remaining Li ions occupy the 96h site . (3) For $3<x<7$, the fractional occupancy in the 24d and 96h sites is linearly interpolated between the occupancy values at the two end-point Li contents. Similar to the case of NASICON, we chose Li/vacnacy configurations with the lowest DFT-relaxed energy among 20 low Ewald energy configurations. With the Li/vacancy configuration determined, we selected the ordering of substituted cations with the lowest Ewald energy. It should be noted that our ordering scheme does not guarantee the generation of thermodynamic ground state cation orderings, and thus all our energy values represent an upper bounds to those of the true ground state structures.

\subsection{DFT and CHGNet relaxations}
DFT calculations were performed within the projector augmented wave (PAW) formalism \cite{Kresse1999_PAW}, using the Vienna ab initio simulation package \cite{Kresse1996_VASP}. All DFT relaxations were performed using Perdew-Burke-Ernzerhof
(PBE) generalized gradient approximation (GGA) functional\cite{Perdew1996_PBE} and converged to $10^{-5}$ eV for energy and 0.05 eV/Å for force respectively. In the CHGNet pre-screening step, we used pretrained CHGNet model \cite{Deng2023_chgnet}. All CHGNet relaxations were converged to 0.1 eV/Å for force.

It should be noted that the Li intercalation voltages predicted by DFT calculations within GGA usually underestimate the experimentally measured voltages\cite{Urban2016_computation}. The discrepancy in the voltage prediction can be reduced by using Hubbard+U method \cite{Zhou2004_Hubbard_Li_intercalation,Isaacs2020_Hubbard_Li_intercalation2,Timrov2022_Hubbard_Li_intercalation3}. In our DFT calculations, the same +U values are applied to the TMs (3d: V, Cr, Mn, Fr, Ni, Co, 4d: Mo, 5d: W) as the default values of MPRelaxSet implemented in \texttt{pymatgen}, where the U values are obtained by fitting to experimental binary formation enthalpies \cite{Wang2006_hubbardU}. 

\subsection{CHGNet fine-tuning and MD simulation}

To run MD simulations for subsets of NASICON and garnet final candidates listed in Tables \ref{tab:subset_candidates_NASICON} and \ref{tab:subset_candidates_garnet}, we fine-tuned two CHGNet models separately for NASICON and garnet compounds, similar to a previous workflow \cite{Jun2024_nitride}. The dataset is constructed by running MD simulations using pretrained CHGNet model, followed by DFT static calculations on selected MD trajectory frames. The MD simulations with pretrained CHGNet model are performed for all compounds listed in Tables \ref{tab:subset_candidates_NASICON} and \ref{tab:subset_candidates_garnet} for a duration of 0.5 ns at multiple temperatures (500 K, 600 K, 700 K, 800 K, 900 K, and 1000 K) with a time-step of 2 fs. We uniformly chose 100 trajectory frames for each MD simulation run, resulting in 10,200 frames and 10,800 frames for NASICON and garnet, respectively. The DFT static calculations for these selected frames are performed with the electronic convergence criteria of $10^{-5}$ eV using DFT settings compatible with the Materials Project database. 

The fine-tuning for garnet CHGNet model is performed by tuning all atom, bond, and angle convolution layers, while keeping all other layers freezed. On the other hand, we found that the same fine-tuning scheme results in unphysical crystal configurations for the NASICON model during the MD simulations. For instance, some tetrahedral \ch{PO4} polyanion groups became trigonal plane \ch{PO3} groups with one broken P-O bond during those MD simulations. Therefore, the fine-tuning for NASICON CHGNet model was performed by freezing all layers, except only the last layers of atom, bond, and angle convolution layers. Both NASICON and garnet CHGNet models are fine-tuned with a 0.9:0.05:0.05 ratio for training, validation, and test datasets. The fine-tuning was conducted up to 100 epochs, and the models with the lowest force error were picked as the optimal models. The energy and force errors of the fine-tuned models are 0 meV and 47 meV/Å for NASICON, and for 2 meV and 37 meV/Å for garnet models.

MD simulations with fine-tuned CHGNet models were performed at various temperatures (500 K, 600 K, 700 K, 800 K, 900 K, and 1000 K). The MD systems were initially heated up to the target temperatures in 2 ps using Berendsen thermodstat, followed by 100 ps of equilibration using Nosé–Hoover thermostat. The ionic trajectories are sampled in the final 1.4-ns production stage using Nosé–Hoover thermostat. All MD simulations were performed in NVT ensemble with a time step of 2 fs. Li-ion conductivity at 300 K and activation energy are obtained by extrapolating from the high-temperature conductivities assuming the Arrhenius relationship. The error error bounds (upper and lower limits) of Li-ion conductivity and the standard deviation of activation energy were evaluated using the scheme previously established \cite{He2018_MD_error}.

\section{Supplemental Information}
\label{sec:SI}
The Supplemental material is available at URL-When-Published.

\setstretch{1}
\bibliography{references}



\end{document}